\documentclass[manuscript,screen]{acmart}

\usepackage{amsmath,amssymb,amsfonts}
\usepackage{algorithmic}
\usepackage{graphicx}
\usepackage{textcomp}
\usepackage{graphicx}
\usepackage{caption}
\usepackage{subcaption}
\usepackage{bbm}
\usepackage{color}
\usepackage{hyperref}
\usepackage{enumitem}
\usepackage{algorithm} 
\usepackage{algorithmic}
\usepackage{enumerate}
\usepackage{booktabs}
\usepackage{stmaryrd}
\usepackage{mathrsfs} 
\usepackage{color}
\usepackage{balance}
\usepackage{adjustbox}
\usepackage{multirow}
\usepackage{svg}

\usepackage[para,online,flushleft]{threeparttable}
\newcommand{\etal}{\textit{et al}. }

\newcommand{\eg}{\textit{e.g.}}

\newcommand{\ie}{\emph{i.e.}}
\AtBeginDocument{%
  \providecommand\BibTeX{{%
    \normalfont B\kern-0.5em{\scshape i\kern-0.25em b}\kern-0.8em\TeX}}}

\setcopyright{acmlicensed}
\copyrightyear{2018}
\acmYear{2018}
\acmDOI{XXXXXXX.XXXXXXX}

\acmConference[Conference acronym 'XX]{Make sure to enter the correct
  conference title from your rights confirmation emai}{June 03--05,
  2018}{Woodstock, NY}
\acmISBN{978-1-4503-XXXX-X/18/06}

\begin{document}

\title[REKI]{Efficient and Deployable Knowledge Infusion for Open-World Recommendations via Large Language Models}

\author{Yunjia Xi}
\email{xiyunjia@sjtu.edu.cn}

\affiliation{%
  \institution{Shanghai Jiao Tong University}
  \city{Shanghai}
  \country{China}
}

\author{Weiwen Liu}
\affiliation{%
  \institution{Huawei Noah's Ark Lab}
  \city{Shenzhen}
  \country{China}}
\email{liuweiwen8@huawei.com}

\author{Jianghao Lin}
\affiliation{%
  \institution{Shanghai Jiao Tong University}
  \city{Shanghai}
  \country{China}
}
\email{chiangel@sjtu.edu.cn}

\author{Muyan Weng}
\affiliation{%
  \institution{Shanghai Jiao Tong University}
  \city{Shanghai}
  \country{China}
}
\email{1355029251@sjtu.edu.cn}

\author{Xiaoling Cai}
\affiliation{%
  \institution{Consumer Business Group, Huawei}
  \city{Shenzhen}
  \country{China}
}
\email{caixiaoling2@huawei.com}

\author{Hong Zhu}
\affiliation{%
  \institution{Consumer Business Group, Huawei}
  \city{Shenzhen}
  \country{China}
}
\email{zhuhong8@huawei.com}

\author{Jieming Zhu}
\affiliation{%
  \institution{Huawei Noah's Ark Lab}
  \city{Shenzhen}
  \country{China}}
\email{jiemingzhu@ieee.org}

\author{Bo Chen}
\affiliation{%
  \institution{Huawei Noah's Ark Lab}
  \city{Shenzhen}
  \country{China}}
\email{chenbo116@huawei.com}

\author{Ruiming Tang}
\affiliation{%
  \institution{Huawei Noah's Ark Lab}
  \city{Shenzhen}
  \country{China}}
\email{tangruiming@huawei.com}

\author{Yong Yu}
\affiliation{%
  \institution{Shanghai Jiao Tong University}
  \city{Shanghai}
  \country{China}
}
\email{yyu@apex.sjtu.edu.cn}

\author{Weinan Zhang}
\affiliation{%
  \institution{Shanghai Jiao Tong University}
  \city{Shanghai}
  \country{China}
}
\email{wnzhang@sjtu.edu.cn}

\renewcommand{\shortauthors}{Yunjia and Weiwen, et al.}

\begin{abstract}
  Recommender system plays a pervasive role in today's online services, yet its closed-loop nature, \ie, training and deploying within a specific closed domain, constrains its access to open-world knowledge. Recently, the emergence of large language models (LLMs) has shown promise in bridging this gap by encoding extensive world knowledge and demonstrating advanced reasoning capabilities. However, previous attempts to directly implement LLMs as recommenders fall short in meeting the demanding requirements of industrial recommender systems, particularly in terms of online inference latency and offline resource efficiency. In this work, we propose an Open-World \underline{R}ecommendation Framework with \underline{E}fficient and Deployable \underline{K}nowledge \underline{I}nfusion from Large Language Models, dubbed \textit{REKI}, to acquire two types of external knowledge about users and items from LLMs. Specifically, we introduce \textit{factorization prompting} to elicit accurate knowledge reasoning on user preferences and items. With factorization prompting, we develop \textit{individual knowledge extraction} and \textit{collective knowledge extraction} tailored for different scales of recommendation scenarios, effectively reducing offline resource consumption. 
  Subsequently, the generated user and item knowledge undergoes efficient transformation and condensation into augmented vectors through a \textit{hybridized expert-integrated network}, ensuring its compatibility with the recommendation task. The obtained vectors can then be directly used to enhance the performance of any conventional recommendation model. We also ensure efficient inference by preprocessing and prestoring the knowledge from the LLM. 
  Extensive experiments demonstrate that REKI significantly outperforms the state-of-the-art baselines and is compatible with a diverse array of recommendation algorithms and tasks. 
  Now, REKI has been deployed to Huawei's news and music recommendation platforms and gained a 7\% and 1.99\% improvement during the online A/B test.
\end{abstract}

\ccsdesc[500]{Information systems~Recommender systems}

\keywords{Recommender Systems; Large Language Models; Knowledge Augmentation}


\maketitle
\section{Introduction}

Recommender systems (RSs) play a pervasive role in contemporary online services, enriching user experiences across diverse domains like music streaming~\cite{van2013deep}, movie exploration~\cite{koren2009matrix}, and e-commerce~\cite{he2016ups}. Nonetheless, one of the main characteristics of recommender systems lies in their \textit{closed-loop} nature -- recommender models are trained and deployed within closed systems, utilizing internally generated data to train and optimize themselves. 
Due to their insulated nature, classical recommender models confine their data and knowledge to specific application domains~\cite{koren2009matrix, DIN} and isolate from the knowledge in the external world, limiting their ability to access a broader range of information and insights. Consequently, this closed-loop nature may increase the possibility of generating outdated, repetitive, and imprecise recommendations~\cite{kong2021review}.  To this end, researchers have revealed that if we expand the knowledge horizon beyond the limited domains, predictive accuracy and generalization capabilities of recommender systems could be significantly boosted~\cite{friedman2023leveraging,lin2023sparks}. In this work, we refer to external world knowledge capable of enhancing recommendations as \textit{open-world knowledge}. Specifically, it typically comprises two categories, namely knowledge augmentation for the two crucial elements in recommendations, users and items. The item knowledge often involves factual information obtainable from the web or induced from its textual description. In contrast, user knowledge is more complex and dynamic, usually entailing a delicate understanding of user preferences and profiles. 

Therefore, we propose a departure from the traditional \textbf{closed-loop systems}. Instead, we advocate for recommender systems to embrace an \textbf{open-world systems} paradigm, actively seeking open-world knowledge from the external world. This shift emphasizes the need to move beyond learning from narrowly defined data, fostering a more dynamic and adaptive recommender system that engages with a broader spectrum of knowledge. 
On the one hand, the inclusion of factual knowledge related to items contributes valuable common-sense information about candidate items, thereby elevating the quality of recommendations. Taking the movie recommendation as an example, the external world it needs contains additional movie features such as \textit{lots, related reports, awards, and critic reviews} which are often absent in the recommendation dataset. This augmentation expands the scope of the original data and enhances the effectiveness of recommendation. On the other hand, knowledge inferred from user behavior history on the user side facilitates a more holistic understanding of users, playing a pivotal role in improving recommendation performance. Unraveling the underlying preferences and motivations that drive user behaviors provides deeper insights and clues about users. Aspects such as personality, occupation, intentions, preferences, and tastes can be reflected in user behavior history~\cite{liu2023joint}. The reasoning of preferences can extend to integrating temporal factors (\eg, holiday-themed movie preferences during Christmas) or responding to external events (\eg, an increased interest in health products during a pandemic). This preference reasoning goes beyond basic behavior patterns identified by classical recommenders, delivering human-like recommendations supported by clear evidence.

Previous efforts have explored injecting external knowledge into recommender systems through methods such as knowledge graphs~\cite{guo2020survey,wang2019knowledge} or cross-domain recommendations~\cite{star,guo2023disentangled,ji2023teachers}. Knowledge graphs are incorporated into the recommender system as side information, explicitly providing rigorously validated external world knowledge to enhance recommendation precision. Cross-domain recommendation methods facilitate the transfer of knowledge from source domains to target domains, thereby strengthening the target domain's recommender system with additional insights. However, constructing comprehensive and accurate knowledge graphs or multi-domain datasets requires considerable extra human effort, and the accessible knowledge remains limited. Moreover, they only focus on extracting factual knowledge from the external world and overlook the reasoning knowledge on user preferences \cite{guo2020survey}. For instance, knowledge graphs used in RS usually focus on complement knowledge for items and rarely consider users~\cite{guo2020survey,wang2019knowledge,wang2019multi} since the user-side knowledge is highly dynamic and challenging to capture within a static knowledge graph.

Recent strides in large language models (LLMs) have brought about revolutions in learning paradigms across various research domains, and they also offer substantial promise in bridging the gap between conventional recommender systems and open-world knowledge~\cite{zhao2023survey,zhu2023large,chen2023large}. Notably, LLMs like GPT-4~\cite{gpt4} and LLaMA~\cite{llama}, owing to their vast model scale and corpus size, have memorized a vast array of knowledge and demonstrated remarkable capabilities encompassing problem-solving, logical reasoning, and creative writing~\cite{bubeck2023sparks}. By learning from an extensive corpus of internet texts, LLMs have successfully encoded a broad spectrum of world knowledge, ranging from basic factual information to intricate societal norms and logical structures~\cite{bubeck2023sparks}. Particularly noteworthy is the proficiency of LLMs in leveraging their encoded knowledge to perform basic logical reasoning that aligns with established facts and relationships~\cite{wei2022chain,zhou2023leasttomost}. This inherent capacity of LLMs to possess factual knowledge about items and exhibit logical reasoning skills is pivotal in bridging the gap between conventional recommender systems and open-world knowledge.
In recent research, numerous studies have explored the application of LLMs for recommendation, which usually transform recommendation tasks and user profiles into prompts~\cite{chatrec,liu2023chatgpt,dai2023uncovering,p5,bao2023tallrec}. It has been observed that LLMs are particularly well-suited for applications such as conversational recommendation~\cite{he2023large,chatrec}, zero-shot/few-shot recommendation~\cite{nir,hou2023large,liu2023chatgpt,lin2023rella,bao2023tallrec}, explainable recommendation~\cite{luo2023unlocking,chatrec}, and cold-start scenarios~\cite{sanner2023large,gong2023unified,lin2023clickprompt}. In the realm of traditional recommendation tasks, such as sequential recommendation and top-N recommendation, the integration of collaborative signals in LLM-based recommendations has indicated the potential to outperform classical recommendation models~\cite{lin2023can,wang2023flip}.

However, in industrial recommendation scenarios characterized by a multitude of users and items and stringent efficiency requirements, directly using LLMs as recommenders still faces significant challenges, which primarily stem from two key shortcomings of deployment:
\begin{itemize}
    \item [1)] \textbf{Online Inference Latency.} To cater to the real-time demands of users, industrial recommendation systems typically have strict latency requirements (usually within 100 milliseconds~\cite{DIR}). However, the inherent limitation of LLMs -- the excessive number of model parameters -- will introduce significant inference latency, making it impractical to directly leverage LLMs as recommenders in industrial settings. With billions of users and thousands of user behaviors, LLMs struggle to meet the low latency requirement of industrial recommendation systems. Moreover, the extensive model size also brings much online computational overhead and hinders the feasibility of incorporating real-time user feedback. 
    \item [2)] \textbf{Offline Resource Consumption.} The term ``resource'' here encompasses both time and computational resources. Drawing insights from prior research~\cite{bao2023tallrec,p5,lin2023rella}, the good performance of LLMs in recommendation scenarios often entails fine-tuning LLMs on specific recommendation data. However, industrial recommendation scenarios typically involve a vast array of users and items, with user behavior distributions evolving over time. Achieving satisfactory results in such scenarios demands continuous fine-tuning of LLMs on regularly updated and substantial amounts of data, consequently leading to significant offline time cost and computational resource consumption. Additionally, fine-tuning LLMs often demands a considerable amount of time. During this period, the online distribution of user behavior may have undergone some changes. This temporal misalignment could result in a situation where a fine-tuned LLM may no longer be suitable for the current online data distribution, leading to inaccuracy recommendations.
\end{itemize}

 Therefore, it is impractical to replace existing recommendation processes and conventional recommendation models (CRMs) with LLMs in industrial recommendations. As a result, we do not engage in fine-tuning LLMs on recommendation data; instead, we focus on efficiently extracting knowledge about users and items from LLMs to augment CRMs. This approach allows us to preserve the advantage of CRMs in terms of inference latency while effectively leveraging open-world knowledge from LLMs as well. Despite the appealing benefits of this approach, efficient extraction and utilization of knowledge from LLMs pose considerable challenges. 
\begin{itemize}
    \item [1)] \textbf{How to extract recommendation-related knowledge from LLMs.} Knowledge related to items may encompass various aspects, while knowledge related to users may involve complex reasoning. When facing complex tasks, LLMs often suffer from the issue of \textit{compositional gap}, where LLMs have difficulty in generating correct answers to the compositional problem like recommending items to users, whereas they can correctly answer all its sub-problems~\cite{press2022measuring}. Therefore, LLMs may also encounter the issue of compositional gap when confronted with knowledge reasoning and extraction encompassing multiple facets, thereby preventing us from fully exploiting the open-world knowledge encoded in LLMs~\cite{kang2023llms,dai2023uncovering}.
     \item [2)] \textbf{How to ensure efficient knowledge extraction in large-scale scenarios.} Although the removal of fine-tuning LLMs has saved considerable time and computational resources, knowledge generation may still be inefficient when dealing with a large volume of users and items. Industrial recommendation scenarios often involve a massive number of users and items, potentially achieving a billion-level. Dealing with such a great number of users and items makes utilizing LLMs to generate knowledge for each user and item a time-consuming and resource-intensive task. 
     \item [3)] \textbf{How to integrate the open-world knowledge into CRMs.} The open-world knowledge generated by LLMs manifests in human-like texts, instead of neural representations that can be directly fed into CRMs. Even in instances where some LLMs are open-sourced, their latent hidden states are typically large dense vectors (\eg, 4096 for each token in LLaMA-7B~\cite{llama}), which are highly abstract and not readily digestible to CRMs. Hence, it is pivotal to effectively transform the output knowledge to align with the recommendation space without loss of information or misinterpretation. Furthermore, due to the hallucination problem~\cite{ji2023survey}, the knowledge produced by LLMs can occasionally be unreliable or misleading, possibly introducing much noise in recommendation.
\end{itemize}
Consequently, it is imperative to extract open-world knowledge from LLMs efficiently while enhancing the accessibility and reliability of the knowledge, which fully unleashes the potential of open-world recommender systems.

To address the above challenges, we propose an Open-World \underline{R}ecommendation Framework with \underline{E}fficient and Deployable \underline{K}nowledge \underline{I}nfusion from LLMs, (dubbed \textit{REKI}\footnote{
This work is an extended version of the paper \href{https://arxiv.org/abs/2306.10933}{Towards Open-World Recommendation with Knowledge Augmented from Large Language Model}, which is accepted by RecSys 24. Compared to the original version, this work emphasizes deployment efficiency, designing collective knowledge for numerous users and items, and includes extensive experiments to compare its advantages.}). REKI is a model-agnostic framework that efficiently bridges classical recommender systems and open-world knowledge. It first leverages LLMs to generate open-world knowledge and then applies CRMs to model collaborative signals and LLM-enhanced knowledge. In this way, we integrate the strengths of both LLMs and CRMs to significantly improve the model's predictive accuracy while concurrently maintaining the low online inference latency of CRMs. 
Specifically, REKI consists of two stages: (1) knowledge extraction and (2) knowledge integration. To avoid the compositional gap for knowledge extraction, we propose \textit{factorization prompting} to break down the complex preference reasoning problem into several vital factors, and then generate the reasoning knowledge on users and the factual knowledge on items. With factorization prompting, we devise two distinct approaches, \textit{individual knowledge extraction} and \textit{collective knowledge extraction}, to ensure efficient knowledge extraction for various scales of recommendation scenarios. The individual knowledge extraction is designed for small-scale scenarios with limited users and items, where LLMs are employed to generate individual knowledge for each user and item. Collective knowledge extraction is designed for large-scale scenarios with massive users and items (\eg, million- or even billion-level), where we first cluster items and users and then generate collective knowledge for each cluster, significantly reducing offline resource consumption.  Then, the knowledge integration stage transforms the generated knowledge into augmented vectors in the recommendation space. In this stage, we propose \textit{hybridized expert integration network} (HEIN) to reduce dimensionality and ensemble multiple experts for robust knowledge learning, thus increasing the reliability and availability of the generated knowledge. Next, the recommendation model incorporates the augmented vectors with original in-domain features for optimization, combining both the in-domain recommendation knowledge and the open-world knowledge. Our main contributions can be summarized as follows:
\begin{itemize}
    \item We present an efficient and deployable open-world recommender system, REKI, which bridges the gap between the in-domain knowledge recommendation and the open-world knowledge from LLMs. To the best of our knowledge, this is the first practical solution that introduces logical reasoning with LLMs for user preferences to the recommendation domain.
    \item REKI transforms the open-world knowledge into dense vectors in recommendation space, compatible with any recommendation task and models. We also release the code of REKI and the generated textual knowledge from LLMs\footnote{Code and knowledge are available at \url{https://github.com/YunjiaXi/REKI} and \url{https://github.com/mindspore-lab/models/tree/master/research/huawei-noah/KAR}} to facilitate future research.
    \item REKI is industrial-friendly and deployable for practical applications. We devise collective knowledge extraction for large-scale scenarios to achieve offline resource efficiency. The knowledge augmentation can be preprocessed and prestored for fast training and inference, avoiding the large inference latency when adopting LLMs to RSs. Now, REKI has been deployed to Huawei's news and music recommendation platforms and gained a 7\% and 1.99\% improvement in the online A/B test. This is one of the first successful attempts at deploying LLM-based recommender to real-world applications.
\end{itemize}
Extensive experiments conducted on public datasets also show that REKI significantly outperforms the state-of-the-art models and is compatible with various recommendation algorithms. We believe that REKI not only sheds light on a way to inject the knowledge from LLMs into the recommendation models, but also provides a practical framework for open-world recommender systems in large-scale applications.

\section{Related Work}
This section reviews studies on recent advances in recommendation with pretrained language models (PLMs).

\subsection{PLM as Recommender Itself}
The emergence of pretrained language models (PLMs) has brought tremendous success in Natural Language Processing (NLP), and PLMs also show great potential in other domains like recommendations~\cite{wu2023survey,wang2023flip}. One promising direction is to leverage PLMs as the primary driver of recommendations, allowing PLMs to directly accomplish the recommendation tasks~\cite{li2023large,wu2023survey,yu2023self}. For example, LMRecSys~\cite{zhang2021language} is one of the earliest attempts to transfer the session-based recommendation task into prompts and evaluate the performance of BERT~\cite{bert} and GPT-2~\cite{gpt2} in movie recommendation. Later, P5~\cite{p5} and M6-Rec~\cite{m6rec} finetune pretrained language models (T5~\cite{t5} or M6~\cite{m6}) by converting multiple recommendation tasks to natural language sequences to incorporate knowledge and semantics inside the training corpora for personalization. Similarly, RecFormer~\cite{li2023text} models user preferences and item features as language representations for the sequential recommendation. In this earlier stage, the sizes of the language models for recommendation are relatively small (\eg, under billions of parameters), and finetuning is usually involved for better performance.

With the scaling of the model size and corpus volume, especially with the emergence of ChatGPT~\cite{gpt4}, LLMs have shown uncanny capability in a wide variety of tasks~\cite{fan2023recommender}. One of the unique abilities of LLMs is reasoning, which emerges only when the model size surpasses a certain threshold. Zero-shot learning or in-context learning is widely used since finetuning LLMs requires lots of resources. Several studies apply LLMs as recommenders and achieve some preliminary results~\cite{kang2023llms,liu2023chatgpt,chatrec,nir,dai2023uncovering,hou2023large}. For instance, ChatRec~\cite{chatrec} employs LLMs as a recommender system interface for conversational multi-round recommendations. Some researchers empirically study conversational recommendation tasks using representative large language models in a zero-shot setting~\cite{he2023large}. Liu \etal~\cite{liu2023chatgpt} investigate whether ChatGPT can serve as a recommender with task-specific prompts and report the zero-shot performance. Hou \etal~\cite{hou2023large} further report the zero-shot ranking performance of LLMs with historical interaction data.

However, directly using LLMs as recommenders generally falls behind state-of-the-art recommendation algorithms, implying the importance of domain knowledge and collaborative signals for recommendation tasks~\cite{kang2023llms,dai2023uncovering,lin2023can}. Therefore, there are also methods exploring the incorporation of recommendation collaborative signals into LLMs through parameter-efficient finetuning approaches~\cite{liu2023pre}. For example, TALLRec~\cite{bao2023tallrec} finetunes LLaMA-7B model~\cite{llama} with a LoRA~\cite{hu2021lora} architecture on recommendation data. Another study~\cite{harte2023leveraging} finetunes an Open-AI ada model\footnote{https://platform.openai.com/docs/guides/fine-tuning} on recommendation data but finds its performance lag behind utilizing the embedding of LLM for similarity matching or as an initialization for recommendation model. Those methods achieve good recommendation performance by finetuning LLMs on recommendation data. Nevertheless, there are a large number of users and items in industrial recommendation scenarios, and the distribution of user behavior tends to change over time. Obtaining satisfactory results on such data usually entails continuous finetuning of LLMs on regularly updated and large amounts of data, leading to significant time consumption and substantial GPU resource usage.

Moreover, these preliminary studies mainly overlook the inference latency during deployment, offline resource consumption, and compositional gap issues of LLMs. In this work, we propose factorization prompting to extract both user and item knowledge from LLMs, alleviating the compositional gap. For large-scale scenarios with massive users and items, we design collective knowledge extraction, which first conducts the user and item clustering and then extracts knowledge for those clusters. This approach significantly reduces the number of knowledge generation, thus improving offline resource efficiency. The extracted knowledge is then adapted to the recommendation domain as augmented representation vectors, which can be prestored for fast training and inference.



\subsection{PLM as Component of Traditional Recommender}
 Unlike the above approaches, where PLMs are used as the primary driver of recommendations, another category of work leverages PLMs as an auxiliary component in traditional RSs. Here, PLMs are usually adopted to encode the textual features (\eg, item descriptions, user reviews) or provide extra knowledge for classical recommendations for better user or item representations~\cite{ubert,zesrec,UniSRec,VQRec}. 
For example, U-BERT~\cite{ubert} leverages the embeddings of user review texts encoded by BERT~\cite{bert} to complement user representations.
ZEREC~\cite{zesrec} incorporates traditional recommender systems with PLMs to generalize from a single training dataset to other unseen testing domains. 
UniSRec~\cite{UniSRec} utilizes BERT~\cite{bert} to encode user behaviors and therefore learns universal sequence representations for downstream recommendation.
Built upon UniSRec, VQ-Rec~\cite{VQRec} further adopts vector quantization techniques to map language embeddings into discrete codes, balancing the semantic knowledge and domain features.
The above methods usually use small PLMs (\eg, BERT-base with 110M parameters) to convert texts to dense vectors, the semantic information of which could be limited and thus fail to provide strong assistance for traditional recommender systems.

Another line of work adopts large language models with billion-level parameters, focusing on encoding or prompting external open-world knowledge from LLMs. For instance, some researchers propose S\&R Multi-Domain Foundation model~\cite{gong2023unified}, which finetunes ChatGLM2-6B~\cite{du2022glm} to extract domain invariant features for promoting search and recommendation performance in cold-start scenarios. Some researchers also find that LLMs provide competitive recommendation performance for pure language-based preferences in the near cold-start case in comparison to item-based CF methods~\cite{sanner2023large}. LLM-Rec~\cite{lyu2023llm} investigates various prompting strategies to generate augmented input text from GPT-3 (\textit{text-davinci-003}), which improves the recommendation capabilities. Another work~\cite{mysore2023large} utilizes InstructGPT (175B)~\cite{ouyang2022training} for authoring synthetic narrative queries from user-item interactions and train retrieval models for narrative-driven recommendations on synthetic data. TagGPT~\cite{li2023taggpt} provides a system-level solution of tag extraction and multi-modal tagging in a zero-shot fashion equipped with GPT-3.5 (\textit{gpt-3.5-turbo}).
Although these methods have made early attempts at utilizing LLMs, they either simply use LLMs as fixed encoders to convert original texts into dense vectors without generating additional textual knowledge, or do not make specific designs to incorporate the generated textual information into traditional recommender systems. This may lead to the instability of RS due to the noise or high dimension of LLM embeddings.

In this paper, we propose factorization
prompting to break down the complex preference reasoning problem into several vital factors to generate open-world knowledge, avoiding the compositional gap. Two knowledge generation approaches, individual and collective knowledge extraction, are introduced to generate knowledge effectively under different scales of recommendation scenarios. Then, we devise a hybridized expert integration network to reduce dimensionality and ensemble multiple experts for robust knowledge learning, thus increasing the reliability and availability of the generated knowledge. The above two stages can be preprocessed and prestored for fast training and inference, which avoids the significant inference latency when using LLMs in RSs. Now, our proposed REKI has been deployed in Huawei news and music recommendation platforms
and has gained significant improvements.


\begin{figure}
    \centering
    \includegraphics[width=\textwidth]{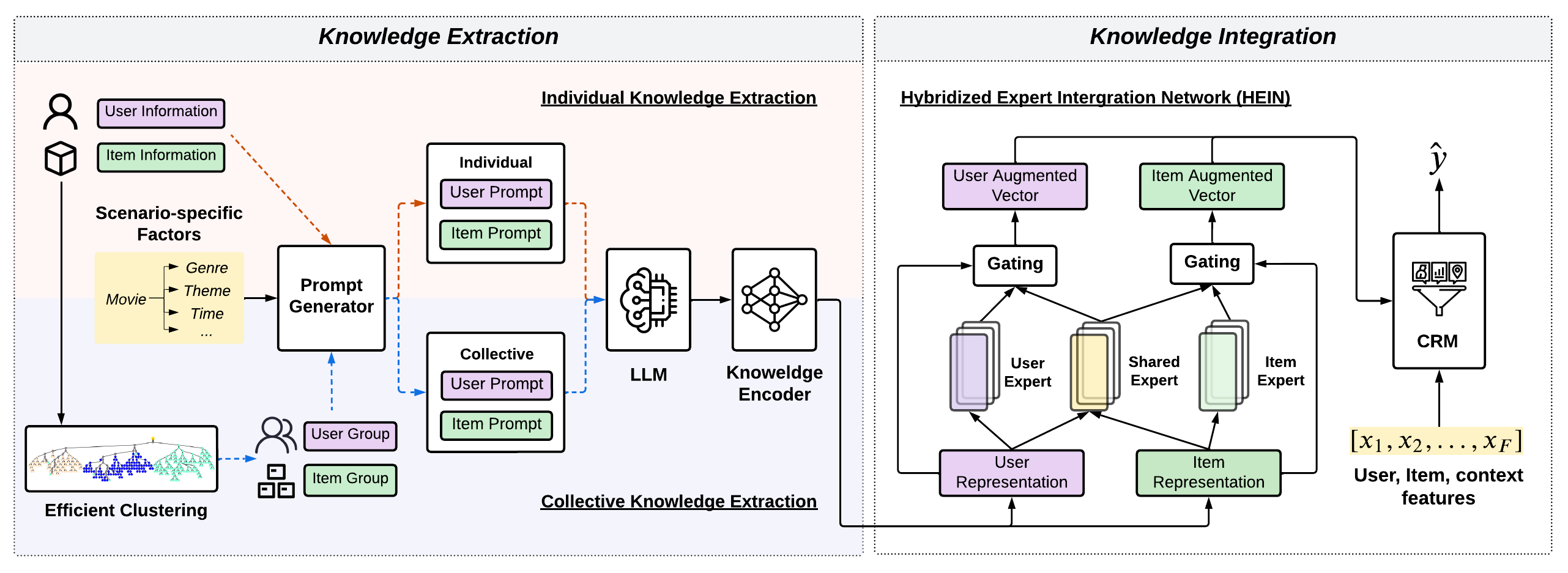}
    \caption{The overall framework of REKI consists of a knowledge extraction stage and a knowledge integration stage. \textbf{Knowledge extraction stage} leverages our designed factorization prompting as well as two kinds of knowledge extraction approaches to extract knowledge from LLMs for users and items effectively. \textbf{Knowledge integration stage} converts the open-world knowledge into compact and user and item representations suitable for recommendation. The user/item representations and user/item augmented vectors can be prestored for fast inference. Next, the user and item augmented vectors are integrated into an existing conventional recommendation model (CRM) as additional input features.}
    \label{fig:open_world}
\end{figure}

\section{Preliminaries}\label{sect:network_structure}
In this section, we formulate the recommendation task and introduce the notations. The core task of recommendation is to precisely estimate the user preference towards each candidate item given a certain context, which can be formulated as a classification problem over multi-field categorical data. The dataset is denoted as $\mathcal D = \{(x_1, y_1, c_1), \ldots, (x_i, y_i, c_i), \ldots, (x_n, y_n, c_n)\}$, where $x_i$ represents the categorical features for the $i$-th instance, $y_i$ denotes the corresponding label, and $c_i$ is the context. Usually, $x_i$ contains sparse one-hot vectors from multiple fields, such as item ID and genre. We can denote the feature of the $i$-th instance as $x_i=[x_{i,1}, x_{i,2},\ldots,x_{i, F}]$ with $F$ being the number of field and $x_{i, k}$, $k=1,\ldots,F$ being the feature of the corresponding field. 

Recommendation models usually aim to learn a function $f(\cdot)$ with parameters $\theta$ that can accurately predict $\hat{y_i} = f(x_i;\theta)$ for each sample $x_i$. In practice, industrial recommender systems are confronted with massive users and items. Thus, their recommendation is usually divided into multiple stages, \ie, candidate generation, ranking, and reranking~\cite{liu2022neural}, where different models are used to narrow down the relevant items.
 
However, these classical recommendation models are typically trained on a specific recommendation dataset (\ie, a closed-loop system), overlooking the potential benefits of accessing open-world knowledge. 

\section{Methodology}
We first provide an overview of our proposed Open-World Recommendation Framework with Efficient and Deployable Knowledge Augmented from Large Language Models, dubbed REKI, and then elaborate on the details of each component.

\subsection{Overview}
As shown in Figure~\ref{fig:open_world}, we design REKI to extract open-world knowledge from LLMs and incorporate it into RSs. This framework is model-agnostic and consists of the following two stages: 

\smallskip
\noindent
\textbf{Knowledge Extraction Stage} leverages our designed factorization prompting to extract recommendation-relevant knowledge from LLMs. We first decompose the complex reasoning tasks by identifying major scenario-specific factors that determine user preferences and item characteristics \eg, theme, genre, and, director. Then, according to different scales of recommendation scenarios, we devise two knowledge generation approaches with those factors. The individual knowledge extraction for the small scenarios utilizes LLMs to generate individual knowledge about the scenario-specific factors for each item and user. In collective knowledge extraction for large scenarios, we first cluster the user and item into groups, respectively, and then require LLMs to generate shared knowledge for each group according to scenario-specific factors. Thus, we can obtain open-world knowledge about the user and item beyond the original recommendation dataset. Finally, textual knowledge obtained from LLMs is encoded into dense representations by a knowledge encoder.

\smallskip
\noindent
\textbf{Knowledge Integration Stage} converts open-world knowledge into compact and relevant representations suitable for recommendations, bridging the gap between LLMs and RSs. First, We devise a hybridized expert integration network (HEIN) to transform the hign-dimension knowledge representations from the semantic space\footnote{the embedding space from language models} to the recommendation space. In this way, we obtain the user augmented vector for user preferences and the item augmented vector for each candidate item. Next, the user and item augmented vectors are integrated into an existing conventional recommendation model (CRM), enabling it to leverage both domain knowledge and open-world knowledge during the recommendation. 

The knowledge extraction stage is conducted through preprocessing. The hybridized expert integration network and CRM are jointly trained in an end-to-end manner.

\subsection{Knowledge Extraction}
As the model size scales up, LLMs can encode a vast array of world knowledge and have shown emergent behaviors such as the reasoning ability~\cite{qiao2023reasoning,huang2022reasoning}. This opens up new possibilities for incorporating user-side knowledge of user preferences and factual knowledge for candidate items in recommender systems. However, it is non-trivial to extract those two kinds of knowledge from LLMs due to the following three challenges.

Firstly, when facing tasks that require the reasoning ability like user preference reasoning~\cite{press2022measuring}, LLMs often suffer from the \textit{compositional gap} where the model fails at generating the correct answer to the compositional question but can correctly answer all its sub-questions. Users' clicks on items are motivated by multiple key aspects, and users' interests are diverse and multifaceted, which involve multiple reasoning steps. To this end, LLMs may not be able to produce accurate reasoning for user preference knowledge directly. Expecting LLMs to provide precise recommendations in one step as in previous works~\cite{liu2023chatgpt,chatrec,hou2023large} might be overly ambitious. 

Secondly, as for factual knowledge about items, LLMs contain massive world knowledge, yet not all of it is useful for recommendation. When the request to an LLM is too general, the generated factual knowledge may be correct but useless, as it may not align with the inferred user preferences. For example, an LLM may infer that a user prefers highly acclaimed movies that have received multiple awards, while the generated factual knowledge is merely about the storyline of the target movie. This mismatch between user-side knowledge and item factual knowledge can limit the LLM-enhanced performance of RSs. 

Lastly, more importantly, the majority of recommendation systems in the industry may face the challenge of dealing with billions of users and items. Given the enormous parameter size of LLMs, the time and resources for generating knowledge can be considerable. Considering the frequent model updates required by industrial recommendation systems and resource constraints, generating knowledge for each user and item becomes impractical in large-scale recommendation scenarios. Additionally, fine-tuning LLMs often demands a considerable amount of time. During this period, the online distribution of user behavior may undergo some changes. This temporal misalignment could result in a situation where a fine-tuned LLM may no longer be suitable for the current online data distribution, leading to inaccuracy recommendations.


Therefore, inspired by the success of Factorization Machines \cite{rendle2010factorization} in RSs, we design \textit{factorization prompting} to first explicitly "factorize" the user preference and item knowledge into several major factors for effectively extracting the open-world knowledge from LLMs. Then, with the factorized factors incorporated into the prompt design, the complex reasoning and knowledge extraction problem can be broken down into simpler subproblems for each factor, thus alleviating the compositional gap of LLMs. In addition, we devise distinct knowledge generation approaches for recommendation scenarios of different scales. For relatively smaller scenarios, we implement individual knowledge extraction, where the item prompt and user prompt are designed for each user and item, facilitating the generation of individual knowledge. In contrast, we introduce collective knowledge extraction for larger scenarios with a considerable number of users and items, which involves clustering users and items first and then prompting collective knowledge for each cluster. Finally,
textual knowledge obtained from LLMs is encoded into dense representations by a knowledge encoder (\ie, open-sourced language models like BERT~\cite{bert}).

\subsubsection{Scenario-specific Factors} 
The factorization prompting first extracts scenario-specific factors that comprehensively cover the critical aspects for both user preferences and item characteristics.
Those factors might vary for different recommendation scenarios. To determine the factors for different scenarios, we rely on a combination of interactive collaboration with LLMs and expert opinions. For example, in movie recommendation, given a prompt ``\textit{List the important factors or features that determine whether a user will be interested in a movie},'' LLMs can provide some potential factors. Then, we involve human experts to confirm and refine the outputs to acquire the final scenario-specific factors for movie recommendation --- including \textit{genre, actors, directors, theme, mood, production quality, and critical acclaim}. Similarly, in news recommendation, we may obtain factors like \textit{topic, source, region, style, freshness, clarity, and impact}. The collaborative process between LLMs and experts ensures that our chosen factors encompass the critical dimensions of user preference and item characteristics for each scenario. 

Note that the specification of these factors is required only once for each scenario and it does not demand much domain expertise with the aid of LLMs. This shows that the proposed method can be easily generalized to different scenarios with little human intervention and manual effort. 
After obtaining the scenario-specific factors, we have designed individual knowledge extraction and collective knowledge extraction for small and large-scale scenarios, which are detailed in Section~\ref{sec:individual} and Section~\ref{sec:collective}, respectively.

\subsubsection{Individual Knowledge Extraction}\label{sec:individual} 
\begin{figure*}
    \centering
    \includegraphics[trim={0.5cm 1cm
    0.5cm 0.3cm},clip,width=\textwidth]{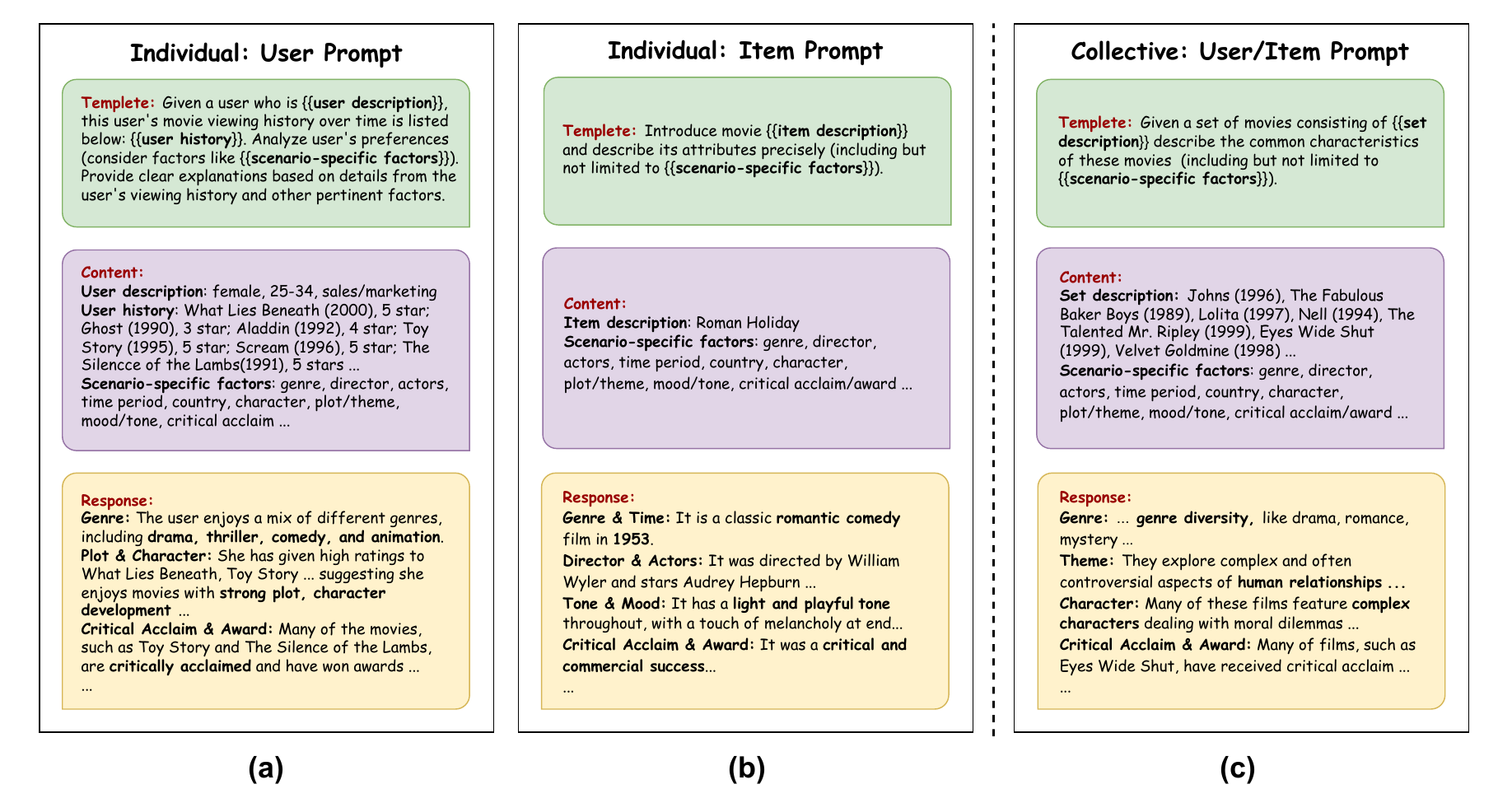}
    \caption{Example prompts for REKI. The green, purple, and yellow text bubbles represent the prompt template, the content to be filled in the template, and the real response generated by LLMs, respectively.}
    \label{fig:promp}
\end{figure*}

Our first prompt engineering approach -- individual knowledge extraction -- is designed explicitly for smaller scenarios where the number of users and items is not substantial. In such scenarios, it is feasible to employ LLMs to generate individual knowledge for each user and item. Specifically, we propose to apply the LLM as a preference reasoner to infer user preferences, and a knowledge provider to acquire external factual knowledge for candidate items. Therefore, we design two types of prompts accordingly: user prompt and item prompt, as illustrated in Figure~\ref{fig:promp}(a) and Figure~\ref{fig:promp}(b).

\textbf{User prompt} is constructed with the user's profile description, behavior history, and scenario-specific factors. Figure~\ref{fig:promp}(a) shows an example of the prompt and real response from the LLM, where the user profile description and behavior history provide the LLM with the necessary context and user-specific information to understand the user's preferences. Scenario-specific factors can instruct the LLM to analyze user preference from different facets and allow the LLM to recall the relevant knowledge more effectively and comprehensively. For example, in the factor of genre, the LLM infers user preference for genres such as thriller, comedy, and animation based on user's positive ratings on thriller movies like \textit{What Lies Beneath}, \textit{Scream}, as well as comedy animations like \textit{Toy Story} and \textit{Aladdin}. With the designed prompt, LLM can successfully analyze the user's preferences toward corresponding factors, which is beneficial for recommendations.


\textbf{Item prompt} is designed to fill the knowledge gap between the candidate items and the generated user-side knowledge. Since the dataset in RS may lack relevant knowledge about scenario-specific factors from items, we need to extract corresponding knowledge from LLM to align the generated user and item knowledge.  As illustrated in Figure~\ref{fig:promp}(b), an item prompt consists of two parts -- the target item description and the scenario-specific factors. This prompt can guide LLM in compensating for the missing knowledge within the dataset. For instance, in Figure~\ref{fig:promp}(b), LLM supplements \textit{Roman Holiday} with \textit{"a light and playful tone"}, which is rarely recorded in the datasets. In this way, LLM provides external knowledge that aligns with user preferences, allowing for more accurate and personalized recommendations.

 By combining the two kinds of prompts, we enable the LLM to act as both a preference reasoner and a knowledge provider, thereby extracting individual user and item knowledge from LLMs. We refer to the generated textual response with the user and item prompt as the \textit{user knowledge} and the \textit{item knowledge}, respectively. This approach of extracting knowledge for each user and item proves effective in scenarios characterized by a relatively limited number of users and items. However, owing to the substantial time and resource consumption involved in generating knowledge through LLMs, the feasibility of extracting knowledge for each user and item becomes questionable when we confront larger recommendation scenarios housing a significant number of users and items (\eg, million-level or even billion-level).

\subsubsection{Collective Knowledge Extraction}\label{sec:collective} For larger recommendation scenarios with a massive number of users and items, we introduce collective knowledge extraction. This approach first conducts user and item clustering, with each cluster represented by a set of items. Subsequently, we have formulated a unified prompt designed for both the user and item to extract knowledge for each cluster in Figure~\ref{fig:promp}(c). Regulating cluster size through clustering algorithms affords us the flexibility to manage the number of clusters, thereby controlling the time and resource consumption during the knowledge generation.

Conventional clustering methods, such as k-means~\cite{ahmed2020k} and standard hierarchical clustering~\cite{murtagh2012algorithms}, often face challenges in handling data within large-scale recommendation scenarios. Firstly, they exhibit high complexity when dealing with large-scale and high-dimension data. Secondly, they lack scalability, struggling to accommodate the continuous arrival of new users or items in dynamic recommendation scenarios. To address these limitations, we choose a streaming hierarchical clustering approach, StreaKHC~\cite{xing2022streaming} algorithm to cluster users and items. Leveraging a scalable point-set kernel, StreaKHC is capable of handling large volumes of streaming data in recommendation scenarios. It employs a top-down search strategy to recursively add a new point to the most similar child node, resulting in a time complexity of $O(n)$ where $n$ is the number of points in the node. Additionally, as a hierarchical clustering algorithm, it forms a tree structure, allowing us to change the number of clusters without regenerating the entire tree.

With this clustering algorithm, we first perform user and item clustering. The clustered results are then processed and incorporated into the user/item prompts, which are fed into the LLM for knowledge extraction. In the following discussion, we will first introduce item clustering and prompts, followed by user clustering and prompts.


\textbf{Item clustering} is a relatively straightforward process. Initially, we pre-train embeddings for all items, item attributes, and user profiles using traditional recommendation models such as DIN~\cite{DIN}. Subsequently, these item and attribute embeddings are then fed into the StreaHKC algorithm for clustering, resulting in multiple clusters. Each item cluster, denoted as $c^a_i$, comprises multiple items represented as $c^a_i=\{a_1,\ldots, a_k,\ldots,a_{m_i}\}$, where 
$a_k$ represents an item, and 
$m_i$ denotes the number of items within that cluster. The maximum value of $m$ can be controlled indirectly through parameters in the StreaHKC algorithm, so we can regulate the number of clusters. The StreaHKC algorithm forms a tree structure. Hence, when some new items arrive, we conduct incremental training to obtain embeddings for these new items. Then, we can efficiently introduce new nodes into this tree and obtain new clusters through a splitting algorithm. 

\textbf{Item prompt} in collective knowledge extraction is designed for extracting the commonalities within a set of items, \ie, collective knowledge. Prompt for item clusters, as depicted in Figure~\ref{fig:promp}(c), comprises set description and scenario-specific factors. Here, the set description refers to each cluster obtained from item clustering, represented by a set of items. Similar to prompts in individual knowledge extraction, the item prompt requires scenario-specific factors to guide LLMs in analyzing the shared characteristics of this set of items. For instance, in Figure~\ref{fig:promp}(c), regarding themes, the LLM can infer that this set of movies "explores complex and often contradictory aspects of human relationships." In terms of characters, "these movies often feature complex characters." Through this prompt, LLMs can effectively reason and extract the commonalities within a set of items, \ie, their collective knowledge. Under the guidance of this prompt, LLM can efficiently generate the item knowledge comprising multiple facets of an item cluster.

\textbf{User clustering} is similar to item clustering, and it utilizes the same embeddings pre-trained mentioned in the item clustering phase. As users exhibit complex and dynamic behavior, we represent users using their click history and profile features. Specifically, for a user $u_k$ with click history $h_k=\{a_1,\ldots,a_k,\ldots,a_{s_i}\}$ of length $s_i$, the average of embeddings for all items recently clicked by the user serves as a part of user representation, while the embedding of user profile features forms the other part. After inputting all the user representations into the StreaKHC algorithm, we obtain each user cluster, $c^u_i=\{u_1,\ldots,u_k,\ldots,u_{t_i}\}$, where $u_k$ represents a user and $t_i$ is the number of users in that cluster. 

\textbf{User prompt} becomes more intricate than item prompt, because it is challenging for untuned LLMs to comprehend users typically identified by IDs without corresponding textual descriptions in recommendation. To this end, we typically represent the user by his or her recent historical behaviors, \ie, a set of items, as we do in the individual knowledge extraction. Here, for a user cluster, we also represent it by a set of items, \ie, the common historical items from all users within this cluster, enabling LLMs to comprehend the cluster for knowledge reasoning and extraction. To achieve this, we extract common historical items from all users within each cluster $c^u_i$, forming a new representation for the user cluster $\hat{c}^u_i$. Mathematically, we first obtain the number of times $N_k$ that each item $a_k$ has been clicked by all users in the cluster $c^u_i$ and then select the top $d$ items with the highest $N_k$ values as new representation for the updated user cluster $\hat{c}^u_i$, \ie,
\begin{equation}   
    \begin{split}
    N_k = &\sum_{u_j} \mathbf{1}_{h_j}(a_k)\,, \forall u_j \in c_i^u, a_k\in \mathcal{I} \\
    \hat{c}^u_i & = \{a_k: \text{Rank}(a_k) \le d\}\,, \\
    \end{split}
\label{eq:cluster}
\end{equation}
where $N_k$ denotes the number of user in cluster $c^u_i$ that have been clicked item $a_k$, $h_j$ is historical clicked item set of user $u_j$, $\mathbf{1}_{h_j}(a_k)$ represents whether the item $a_k$ appears in the click history $h_j$, and $\mathcal{I}$ is the whole item set. Function $\text{Rank}(a_k)$ denotes the position of $a_k$ when all the items $a_k\in \mathcal{I}$ are arranged by descending order of $N_k$. In this way, we have successfully represented a user cluster with a set of items. In other words, the format of our user cluster is now identical to that of the previous item cluster. Therefore, we can extract user collective knowledge from the user cluster using a similar prompt and process with the item prompt, as illustrated in Figure~\ref{fig:promp}(c).

\subsubsection{Knowledge Encoder} The knowledge generated by LLMs is usually in the form of text, which cannot be directly leveraged by traditional RSs that typically process categorical features. Therefore, we design the knowledge encoder module to encode the generated textual knowledge into dense vectors and aggregate them effectively. To harness the potential of textual knowledge generated by LLMs, we employ encoders like BERT~\cite{bert} and ChatGLM~\cite{du2022glm} to obtain the encodings for each token within the text. Then, we require an aggregation process that combines each token to generate the \textit{user representation} $r^u_i\in\mathbb{R}^{m}$ and the \textit{item  representation} $r^\iota_i\in\mathbb{R}^{m}$ of size $m$ as follows
\begin{equation}
    \begin{split}       
    &r^p_i = \text{Aggr}(\text{Encoder}(klg^u_i))\,, \\
    &r^\iota_i = \text{Aggr}(\text{Encoder}(klg^\iota_i))\,,
    \end{split}
\label{eq:aggr_encoder}
\end{equation}
where $klg^u_i$ and $klg^\iota_i$ denote the user knowledge and item knowledge generated by LLMs of the $i$-th instance in the dataset. Here, various aggregation functions can be employed, such as the representation of the [CLS] token and average pooling. In practice, we primarily adopt average pooling. Note that the knowledge encoder is devised for situations where we only have access to the textual outputs of LLMs via API calls. The separate knowledge encoder can be eliminated if we adopt open-sourced LLMs (\eg, LLaMA~\cite{llama}) and the dense hidden states are available.

\subsection{Knowledge Integration}
The knowledge generated by LLMs presents new challenges in harnessing its potential to assist recommendation models: 1) The knowledge representations obtained in the preceding stage are typically large dense vectors (such as 4096 when taking ChatGLM~\cite{du2022glm} as the encoder), and they lie in a semantic space which is significantly different from the recommendation space. Utilizing such vectors directly in CRMs may lead to training instability. Therefore, it is pivotal to effectively transform the output knowledge to align with the recommendation space without loss of information or misinterpretation for the quality of recommendations. 2) Another challenge is that not all knowledge generated by LLMs is reliable. LLMs may suffer from hallucination problem~\cite{ji2023survey}, resulting in noise or unreliable information. Without careful handling, there is a risk that this noise information may be introduced into the CRMs, thereby causing a decline in recommendation accuracy.

To address these challenges, we have devised  \textit{hybridized expert integration network} (HEIN). The HEIN converts dense vectors from the semantic space to the recommendation space. It introduces multiple sets of experts to perform dimensionality reduction and space transformation on knowledge representation $r_i^p$ and $r_i^l$. Using gating mechanisms to combine the results from multiple experts enhances the stability and robustness of the transformation, thereby reducing the impact of noise information on CRMs. Thus, HEIN increases the reliability and availability of generated knowledge and bridges the gap between LLMs and RSs. Then, the output of HEIN will be integrated into a CRM as additional input features, enabling it to leverage open-world knowledge.



\subsubsection{Hybridized Expert Integration Network} To effectively transform and compact the attained aggregated representations from the semantic space to the recommendation space, we propose a hybridized expert integration network (HEIN). The aggregated representations capture diverse knowledge from multiple aspects, so we employ a structure that mixes shared and dedicated experts, inspired by the Mixture of Experts (MoE)~\cite{MoE} approach. This allows us to fuse knowledge from different facets and benefit from the inherent robustness offered by multiple experts. 

In particular, to fully exploit the shared information of the user representation and the item representation, we have designed both shared experts and dedicated experts for each kind of representation. 
The shared experts capture the common aspects, such as shared features, patterns, or concepts, that are relevant to both user and item knowledge. User and item representations also have their dedicated sets of experts to capture the unique characteristics specific to the user or item knowledge. 
Mathematically, denote $\mathcal S_s$, $\mathcal S_u$, and $\mathcal S_\iota$ as the sets of shared experts and dedicated experts for user and item knowledge with the expert number of $n_s$, $n_u$ and $n_\iota$. The output is the \textit{user augmented vector} $\hat{r}_{i}^u\in\mathbb{R}^{q}$ and the \textit{item augmented vector} $\hat{r}_{i}^\iota\in\mathbb{R}^{q}$ of size $q$ ($q$ is much less than the original dimension $m$), which are calculated as follows
\begin{equation}
\begin{split}
    \alpha^u_{i} &=\operatorname{Softmax}(g^u(r^u_i))\,,\quad\alpha_{i}^\iota =\operatorname{Softmax}(g^\iota(r^\iota_i))\,,\\
    \hat{r}_{i}^u&=\sum\nolimits_{e\in \mathcal S_s} \alpha_{i,e}^u \times e(r^u_i) + \sum\nolimits_{e\in \mathcal S_u} \alpha_{i,e}^u \times e(r^u_i)\,,\\
    \hat{r}_{i}^\iota&=\sum\nolimits_{e\in \mathcal S_s} \alpha_{i,e}^\iota \times e(r^\iota_i) + \sum\nolimits_{e\in \mathcal S_\iota} \alpha_{i,e}^\iota \times e(r^\iota_i)\,,\\
\end{split}
\end{equation}
where $g^u(\cdot)$ and $g^\iota(\cdot)$ are the gating networks for user and item representations, and their outputs $\alpha^u_i$ and $\alpha^\iota_i$ are of size $n_s+n_u$ and $n_s+n_\iota$. Here $e(\cdot)$ denotes the expert network, and $\alpha_{i,e}^u$ and $\alpha_{i,e}^\iota$ are the weights of expert $e(\cdot)$ generated by the gating network for user and item, respectively. Here, each expert network $e(\cdot)$ is designed as a Multi-Layer Perceptron (MLP), facilitating dimensionality reduction and space transformation. Empirical experiments show that about 5 experts are sufficient to generate a satisfying performance\footnote{There is no need to set a large number of experts here, and thus, the model does not suffer from convergence issues.}.

\subsubsection{Knowledge Utilization}
Once we have obtained the user augmented vector and the item augmented vector, we can then incorporate them into backbone recommendation models. This section explores a straightforward approach where these augmented vectors are directly treated as \textbf{additional input features}. Specifically, we use them as additional feature fields in recommendation models, allowing them to interact with other features explicitly. During training, the HEIN module is jointly optimized with the backbone model to ensure the transformation process adapts to the current data distribution. Generally, REKI can be formulated as
\begin{equation}
    \hat{y_i} = f(x_i,h_i,c_i,\hat{r}_{i}^u,\hat{r}_{i}^\iota;\theta)\,,
     \label{eq:ctr-goal-3}
\end{equation}
which is enhanced by the the user augmented vector $\hat{r}_{i}^u$ and the item augmented vector $\hat{r}_{i}^\iota$. Importantly, REKI only modifies the input of the backbone model and is independent of the design and loss function of the backbone model, so it is flexible and compatible with various backbone model designs. Furthermore, it can be extended to various recommendation tasks, such as sequential recommendation and direct recommendation, by simply adding two augmented vectors in the input. By incorporating the knowledge augmented vectors, REKI combines both the open-world knowledge and the recommendation domain knowledge in a unified manner to provide more informed and personalized recommendations.


\subsection{Speed-up Approaches for Online Inference}\label{sec:speed_up}
Our proposed REKI framework adopts LLMs to generate knowledge for users and items. Due to the immense scale of the model parameters, the inference of LLMs takes extensive computation time and resources, and the inference time may not meet the latency requirement in real-world recommender systems with large user and item sets. 

To address this, we employ an acceleration strategy to \textbf{prestore knowledge representations} $r^u_i$ and $r^\iota_i$ generated by the knowledge encoder or the LLM into a database. As such, we only use the LLM and knowledge encoder once before training backbone models. During the training and inference of the backbone model, relevant representations are retrieved from the database. Besides, the efficiency of offline knowledge generation with LLM can further be enhanced via quantization or hardware acceleration techniques.

If we have stricter requirements for inference time or storage efficiency, we can detach the HEIN module from the model after training and further \textbf{prestore augmented vectors} \ie, $\hat{r}_{i}^u$ and $\hat{r}_{i}^\iota$, for inference. The dimension of the augmented vectors (\eg, 32) is usually much smaller than that of the knowledge representations (\eg, 4096), which improves the storage efficiency. Additionally, prestoring the augmented vectors reduces the inference time to nearly the same as the original backbone model, and we have provided experimental verification in Section~\ref{sec:speed_up_efficiency}. In particular, assume the inference time complexity of the backbone model is $O(f(n, m))$, where $n$ is the number of fields and $m$ is the embedding size. The polynomial function $f(n,m)$ varies depending on different backbone models. With REKI, the inference time complexity is $O(f(n+2,m))=O(f(n,m))$, which is equivalent to the complexity of the original model. 


Since item features are relatively fixed and do not change frequently, it is natural and feasible to prestore the item knowledge for further use. Moreover, user behaviors evolve over time, making it challenging for LLMs to provide real-time user-side knowledge about behaviors. However, considering that long-term user preferences are relatively stable, and the backbone model already emphasizes modeling recent user behaviors, it is unnecessary to require LLMs to have access to real-time behaviors. Therefore, LLM can infer long-term preferences based on users' long-term behaviors, allowing for conveniently prestoring the generated knowledge without frequent updates. As such, the inference overhead of LLMs can also be significantly reduced. The backbone models can capture ever-changing short-term preferences with timely model updates.  This can take better advantage of both LLMs and recommendation models. Similar to common practice for cold start users or items~\cite{zhang2014addressing}, we use default vectors when encountering new users or items at inference time. Subsequently, we will generate the knowledge for those new users and items offline and add them to the database.

\vspace{-5pt}
\section{Experiment}
In this section, we aim to address the following research questions (RQs) and comprehensively evaluate the performance and versatility of REKI.
\begin{itemize}
    \item \textbf{RQ1:} What improvements can REKI bring to backbone models on different tasks, such as CTR prediction and reranking? 
    \item \textbf{RQ2:} How does REKI perform compared with other PLM-based baseline methods?
    \item \textbf{RQ3:} Does the knowledge from LLM outperform other knowledge-based methods, such as the knowledge graph?
    \item \textbf{RQ4:} Does REKI gain performance improvement when deployed online?
    \item \textbf{RQ5:} Does the acceleration strategy and collective knowledge extraction enhance the online inference and offline resource efficiency?
    \item \textbf{RQ6:} How do the user knowledge and item knowledge generated by the LLM, as well as different knowledge integration approaches, contribute to performance improvement?  
\end{itemize}

\vspace{-4pt}
\subsection{Experimental Settings}
\subsubsection{Datasets}
Our experiments are conducted on two public datasets, MovieLens-1M\footnote{\url{https://grouplens.org/datasets/movielens/1m/}} and Amazon-Book\footnote{\url{https://cseweb.ucsd.edu/~jmcauley/datasets/amazon_v2/}}. 
\begin{itemize}
    \item \textbf{MovieLens-1M} contains 1 million ratings provided by 6000 users for 4000 movies. Following the data processing similar to DIN~\cite{DIN}, we convert the ratings into binary labels by labeling ratings of 4 and 5 as positive and the rest as negative. The data is split into training and testing sets based on user IDs, with 90\% assigned to the training set and 10\% to the testing set. The dataset contains user features like age, gender, occupation, and item features like item ID and category. The inputs to the models are user features, user behavior history (the sequence of viewed movies with their ID, category, and corresponding ratings), and target item features. 
    \item \textbf{Amazon-Book}~\cite{ni2019justifying} is the “Books” category of the Amazon Review Dataset. After filtering out the less-interacted users and items, we remain 11,906 users and 17,332 items with 1,406,582 interactions. The preprocessing is similar to MovieLens-1M, with the difference being the absence of user features. Additionally, ratings of 5 are regarded as positive and the rest as negative.
\end{itemize}
The detailed statistics of the processed datasets are shown in Table \ref{tab:dataset_stats}.

\begin{table}[]
\caption{The preprocessed dataset statistics.}
\begin{tabular}{cccc}
\toprule
\textbf{Dataset} & \textbf{\#user} & \textbf{\#item} & \textbf{\#interaction} \\
\midrule
MovieLens-1M     & 6000            & 4000            & 1,000,209              \\
Amazon-Books     & 11,906          & 17,332          & 1,406,582             \\
\bottomrule
\end{tabular}
\label{tab:dataset_stats}
\end{table}


\subsubsection{Backbone Models} Because REKI is a model-agnostic framework, various tasks and models in traditional ID-based recommendation can serve as the backbone model, taking as input the knowledge-augmented vectors generated by REKI. Here, we select two crucial recommendation tasks: \textit{CTR prediction} and \textit{reranking}, to validate the effectiveness of REKI across various tasks. 

\noindent
\textbf{CTR Prediction.} CTR prediction aims to anticipate how likely a user is to click on an item, usually used in the ranking stage of recommendation. We choose 9 representative CTR models as our backbone models, which can be categorized into user behavior models and feature interaction models. 

\textit{User Behavior Models} emphasize modeling sequential dependencies of user behaviors. 
\begin{itemize}
    \item \textbf{DIN}~\cite{DIN} utilizes attention to model user interests dynamically with respect to a certain item. 
    \item \textbf{DIEN}~\cite{DIEN} extends DIN by introducing an interest evolving mechanism to capture the dynamic evolution of user interests over time.
\end{itemize}
 
\textit{Feature Interaction Models} focus on modeling feature interactions between different feature fields. 
\begin{itemize}
    \item \textbf{DeepFM}~\cite{DeepFM} is a classic CTR model that combines factorization machine (FM) and neural network to capture low-order and high-order feature interactions. 
    \item \textbf{xDeepFM} \cite{xDeepFM} leverages the power of both deep network and Compressed Interaction Network to generate feature interactions at the vector-wise level. 
    \item \textbf{DCN}~\cite{DCN} incorporates cross-network architecture and the DNN model to learn the bounded-degree feature interactions. \item \textbf{DCNv2}~\cite{DCNv2} is an improved framework of DCN which is more practical in large-scale industrial settings. 
    \item \textbf{FiBiNet}~\cite{FiBiNET} can dynamically learn the feature importance by Squeeze-Excitation network and fine-grained feature interactions by bilinear function. 
    \item \textbf{FiGNN}~\cite{Fi-GNN} converts feature interactions into modeling node interactions on the graph for modeling feature interactions in an explicit way. 
    \item \textbf{AutoInt}~\cite{AutoInt} adopts a self-attentive neural network with residual connections to model the feature interactions explicitly.
\end{itemize}

\noindent
\textbf{Reranking.}
Reranking reorders the items from the previous ranking stage by modeling the list-wise influence and derives a list that yields more utility and user satisfaction~\cite{liu2022neural}. We implement the following 4 state-of-the-art reranking models as backbone models. 
\begin{itemize}
    \item \textbf{DLCM} ~\cite{ai2018learning} first applies GRU to encode and rerank the top results.
    \item \textbf{PRM}~\cite{pei2019personalized} employs self-attention to model the mutual influence between any pair of items and users' preferences. 
    \item \textbf{SetRank}~\cite{pang2020setrank} learns permutation-equivariant representations for the inputted items via self-attention. 
    \item \textbf{MIR} \cite{xi2022multi} models the set-to-list interactions between candidate set and history list with personalized long-short term interests.
\end{itemize} 

\subsubsection{PLM-based Baselines}
As for baselines, we compare REKI with methods that leverage pretrained language model to enhance recommendation, such as P5~\cite{p5}, UniSRec~\cite{UniSRec}, VQRec~\cite{VQRec}, TALLRec~\cite{bao2023tallrec}, and LLM2DIN~\cite{harte2023leveraging}. 
\begin{itemize}
    \item \textbf{P5}~\cite{p5} is a text-to-text paradigm that unifies recommendation tasks and learns different tasks with the same language modeling objective during pretraining. 
    \item \textbf{UniSRec}~\cite{UniSRec} designs a universal sequence representation learning approach for sequential recommenders, which introduces contrastive pretraining tasks to effective transfer across scenarios. 
    \item \textbf{VQ-Rec}~\cite{VQRec} uses Vector-Quantized item representations and a text-to-code-to-representation scheme, achieving effective cross-domain and cross-platform sequential recommendation. 
    \item \textbf{TALLRec}~\cite{bao2023tallrec} finetunes LLaMa-7B~\cite{llama} with a LoRA architecture on recommendation tasks and enhances the recommendation capabilities of LLMs in few-shot scenarios. In our experiment, we implement TALLRec with LLaMa-2-7B-chat\footnote{https://huggingface.co/meta-llama/Llama-2-7b-chat-hf}, since it has better performance and ability of instruction following. 
    \item \textbf{LLM2DIN} initializes DIN with item embeddings obtained from chatGLM\cite{du2022glm}, following \cite{harte2023leveraging}.
\end{itemize}
We utilize the publicly available code of these three models and adapt the model to the CTR task with necessary minor modifications. We also align the data and features for all the methods to ensure fair comparisons.

\subsubsection{Evaluation Metrics} On the CTR prediction task, we employ widely-used \textit{AUC} (Area under the ROC curve) and \textit{LogLoss} (binary cross-entropy loss) as evaluation metrics following~\cite{DCNv2,AutoInt,DeepFM,DIN}. A higher AUC value or a lower Logloss value, even by a small margin (\eg, 0.001), can be viewed as a significant improvement in CTR prediction performance, as indicated by previous studies~\cite{xDeepFM,DCNv2,lin2023map}. As for reranking task, several widely used metrics, \textit{NDCG@K}~\cite{ndcg} and \textit{MAP@K}~\cite{yue2007support}, are adopted, following previous work~\cite{xi2022multi,ai2018learning,pei2019personalized}.

\subsubsection{Implementation Details} 
We utilize a widely-used LLM API to generate the knowledge we need.
During our experiment, we utilized approximately 33,000 knowledge messages\footnote{Code and knowledge are available at  \url{https://github.com/YunjiaXi/REKI} and \url{https://github.com/mindspore-lab/models/tree/master/research/huawei-noah/KAR}}, with an average token length of around 550. And the total cost is about \$150, including some preliminary exploratory experiments. 
In the previous methods, we have designed two approaches for knowledge generation: generating individual knowledge for each item and user (referred to as \textbf{REKI-I}  in experiments) and clustering items and users before generating collective knowledge for each cluster (referred to as \textbf{REKI-C} in experiments). Apart from the difference in knowledge generation, the encoders for REKI-I and REKI-C in the experimental section are also different. Unless explicitly stated, REKI-I employs ChatGLM-6B as the encoder, while REKI-C uses BERT as the encoder. We will explain the reason for this choice and compare the performance of two variants in Section~\ref{sec:infer_compar}. Besides, to compensate for the lack of individual information in REKI-C, we also utilize the same encoder to encode the recent history of each user and the descriptions of items as additional augmented vectors. If not specified, REKI generally represents REKI-I.
Then, after the encoder, the average pooling is utilized as the aggregation function in Eq.~\eqref{eq:aggr_encoder}. Each expert in HEIN is implemented as an MLP with a hidden layer size of [128, 32]. The number of experts varies slightly across different backbone models, typically with 1-5 shared experts and 1-6 dedicated experts. We keep the embedding size of the backbone model as 32, and the output layer MLP size as [200, 80]. For the item clustering in REKI-C, we set the maximum number of items in each cluster to 80. For user clustering, the maximum number of users per cluster is set to 100 and 150 on MovieLens-1M and Amazon-Books datasets, respectively. When selecting common historical items for user clustering, we set the threshold $d$ in Eq.~\eqref{eq:cluster} to 15. Other parameters, such as batch size and learning rate, are determined through grid search to achieve the best results. For fair comparisons, the parameters of the backbone model and the baselines are also tuned to achieve their optimal performance.

\begin{table*}[t]
\centering
\caption{Comparison between REKI and backbone CTR prediction models. Note that REKI-I and REKI-C both demonstrate significant improvements over the backbone CTR prediction models (t-test with $p$-value $<$ 0.05).}
\scalebox{0.88}{
\setlength{\tabcolsep}{1.3mm}{
\begin{tabular}{ccccccc|cccccc}
\toprule
\multirow{3}{*}{\textbf{\begin{tabular}[c]{@{}c@{}}Backbone\\ model\end{tabular}}} & \multicolumn{6}{c|}{\textbf{MovieLens-1M}} & \multicolumn{6}{c}{\textbf{Amazon-Books}} \\
\cmidrule{2-13}
 & \multicolumn{3}{c}{\textbf{AUC}} & \multicolumn{3}{c|}{\textbf{Logloss}} & \multicolumn{3}{c}{\textbf{AUC}} & \multicolumn{3}{c}{\textbf{Logloss}} \\
  \cmidrule{2-13}
 & \textbf{base} & \textbf{REKI-C} & \textbf{REKI-I} & \textbf{base} & \textbf{REKI-C} & \textbf{REKI-I} & \textbf{base} & \textbf{REKI-C} & \textbf{REKI-I} & \textbf{base} & \textbf{REKI-C} & \textbf{REKI-I} \\
 \midrule
DCNv2 & 0.7830 & 0.7918 & \textbf{0.7935} & 0.5516 & 0.5432 & \textbf{0.5410} & 0.8269 & 0.8345 & \textbf{0.8350} & 0.4973 & 0.4871 & \textbf{0.4865} \\
DCNv1 & 0.7828 & 0.7918 & \textbf{0.7931} & 0.5528 & 0.5428 & \textbf{0.5411} & 0.8268 & 0.8342 & \textbf{0.8348} & 0.4973 & 0.4878 & \textbf{0.4869} \\
DeepFM & 0.7824 & \textbf{0.7920} & 0.7919 & 0.5518 & 0.5429 & \textbf{0.5432} & 0.8269 & 0.8341 & \textbf{0.8347} & 0.4969 & 0.4889 & \textbf{0.4873} \\
FiBiNet & 0.7820 & 0.7925 & \textbf{0.7938} & 0.5531 & 0.5422 & \textbf{0.5405} & 0.8269 & 0.8342 & \textbf{0.8351} & 0.4973 & 0.4882 & \textbf{0.4870} \\
AutoInt & 0.7821 & 0.7918 & \textbf{0.7935} & 0.5520 & 0.5427 & \textbf{0.5405} & 0.8262 & 0.8344 & \textbf{0.8357} & 0.4981 & 0.4873 & \textbf{0.4863} \\
FiGNN & 0.7832 & 0.7923 & \textbf{0.7935} & 0.5510 & \textbf{0.5431} & 0.5437 & 0.8270 & 0.8345 & \textbf{0.8352} & 0.4977 & 0.4888 & \textbf{0.4870} \\
xDeepFM & 0.7823 & 0.7924 & \textbf{0.7934} & 0.5520 & 0.5430 & \textbf{0.5428} & 0.8271 & 0.8344 & \textbf{0.8351} & 0.4971 & 0.4869 & \textbf{0.4866} \\
DIEN & 0.7853 & 0.7944 & \textbf{0.7953} & 0.5494 & 0.5395 & \textbf{0.5394} & 0.8307 & 0.8374 & \textbf{0.8391} & 0.4926 & 0.4851 & \textbf{0.4812} \\
DIN & 0.7863 & 0.7952 & \textbf{0.7961} & 0.5486 & 0.5398 & \textbf{0.5370} & 0.8304 & 0.8396 & \textbf{0.8418} & 0.4937 & 0.4808 & \textbf{0.4801} \\
\bottomrule
\end{tabular}
}
}
\label{tab:backbone_model}
\end{table*}
\subsection{Effectiveness Comparison}
\subsubsection{Improvement over Backbone CTR Prediction Models (RQ1)} 

On the CTR prediction task, we implement our proposed REKI-I and REKI-C upon 9 representative CTR models, and the results are shown in Table~\ref{tab:backbone_model}. From the table, we can have the following observations: 

First, as a model-agnostic framework, REKI can be applied to various types of baseline models, whether focusing on feature interaction or behavior modeling, and it significantly improves the performance of backbone CTR models on both datasets. For example, when using the feature interaction method FiBiNet as the backbone model on MovieLens-1M, REKI-I achieves a 1.49\% increase in AUC and a 2.27\% decrease in LogLoss. On the Amazon-Books dataset, applying REKI-I with the behavior modeling approach DIN as the backbone model resulted in a 1.38\% improvement in AUC and a 2.77\% improvement in Logloss. Those improvements demonstrate the effectiveness of incorporating open-world knowledge from LLMs into RSs. Moreover, with the equipment of REKI-I and REKI-C, the selected 9 representative CTR models on two datasets all achieve an AUC improvement of about 1-1.5\%, indicating the universality of the REKI. 



Second, REKI-I and REKI-C show more remarkable improvement in feature interaction models compared to user behavior models. Taking MovieLens-1M as an example, REKI-I exhibited improvements in the user modeling model DIN, with a 1.24\% increase in AUC, while in the feature interaction model FiBiNet, it achieved a 1.49\% improvement in AUC. Notably, REKI demonstrated greater enhancements when applied to the feature interaction model. This may be because the knowledge augmented vectors generated by REKI are utilized more effectively by the feature interaction layer than the user behavior modeling layer. The dedicated feature interaction design may better exploit the information contained in the knowledge vectors. 

Generally, in CTR tasks, REKI-C exhibits a performance decrease of 0.1\%-0.2\% compared to REKI-I. This is because using REKI-C with clustered collective knowledge may result in some information loss compared to individual knowledge generated for each user. However, relative to REKI-I, REKI-C can significantly reduce the time required for the knowledge generation, training, and inference time of the backbone model, as presented in Section~\ref{sec:infer_compar}. Therefore, considering the substantial improvement in efficiency, such a level of accuracy loss is acceptable. The detailed efficiency analysis will also be provided in Section~\ref{sec:efficiency}.

\subsubsection{Improvement over Backbone Reranking Models (RQ1)} 
We also incorporate REKI into the state-of-the-art reranking models. The results on MovieLens-1M and Amazon-Books dataset are presented in Table~\ref{tab:reranking-ml} and Table~\ref{tab:reranking-ab} respectively, from which the following observations can be made: 
\begin{itemize}
    \item [(i)] REKI significantly enhances the performance of backbone reranking models. For example, when PRM is employed as the backbone, REKI-I achieves a remarkable increase of 5.71\% and 4.71\% in MAP@7 and NDCG@7 on the Amazon-Books dataset. This indicates that the open-world knowledge extracted by REKI from LLMs can also be applied to reranking tasks, improving the performance of reranking.
    \item [(ii)] REKI demonstrates more pronounced improvements in methods that do not involve history modeling, such as DLCM, PRM, and SetRank. The user preference knowledge provided by REKI is particularly advantageous for these methods. However, the enhancement is slightly smaller in MIR, which adequately explores the relationship between history and candidates. This may be because REKI's reasoning of user preferences and the reranking model's modeling of historical behavior have some overlap, thus providing a more significant enhancement for methods that lack history modeling. 
    \item [(iii)] The performance of REKI-C and REKI-I on reranking tasks is similar to that on CTR tasks, with REKI-I generally performing slightly better than REKI-C. However, this effect is also influenced by the backbone model and may exhibit some exceptions. For example, when using DLCM as the backbone model, REKI-C performs better than REKI-I on the MovieLens-1M dataset. This indicates that REKI-C can achieve performance similar to REKI-I in reranking tasks while efficiently completing knowledge extraction.
\end{itemize}

\begin{table*}[h]
\centering
\caption{The comparison of REKI and backbone reranking models on MovieLens-1M dataset. Note that REKI-I and REKI-C both demonstrate significant improvements over the backbone rerankin models (t-test with $p$-value $<$ 0.05).}
\scalebox{0.88}{
\setlength{\tabcolsep}{1.3mm}{
\begin{tabular}{ccccccc|cccccc}
\toprule
\multirow{2}{*}{\textbf{\begin{tabular}[c]{@{}c@{}}Backbone\\ Model\end{tabular}}} & \multicolumn{3}{c}{\textbf{MAP@3}} & \multicolumn{3}{c|}{\textbf{MAP@7}} & \multicolumn{3}{c}
{\textbf{NDCG@3}} & \multicolumn{3}{c}{\textbf{NDCG@7}} \\
\cmidrule{2-13}
 & \textbf{base} & \textbf{REKI-C} & \textbf{REKI-I} & \textbf{base} & \textbf{REKI-C} & \textbf{REKI-I} & \textbf{base} & \textbf{REKI-C} & \textbf{REKI-I} & \textbf{base} & \textbf{REKI-C} & \textbf{REKI-I} \\
 \midrule
DLCM & 0.7328 & \textbf{0.7508} & 0.7506 & 0.6899 & \textbf{0.7126} & 0.7100 & 0.7176 & \textbf{0.7353} & 0.7339 & 0.7766 & \textbf{0.7902} & 0.7899 \\
PRM & 0.7432 & 0.7521 & \textbf{0.7561} & 0.7029 & 0.714 & \textbf{0.7202} & 0.7285 & 0.7373 & \textbf{0.7418} & 0.7848 & 0.7923 & \textbf{0.7952} \\
SetRank & 0.7424 & 0.7554 & \textbf{0.7584} & 0.7037 & 0.7196 & \textbf{0.7238} & 0.7283 & 0.741 & \textbf{0.7443} & 0.7844 & 0.7943 & \textbf{0.7972} \\
MIR & 0.7629 & 0.7687 & \textbf{0.7696} & 0.7281 & 0.736 & \textbf{0.7372} & 0.7482 & 0.7551 & \textbf{0.7558} & 0.8006 & 0.8057 & \textbf{0.8066}\\
\bottomrule
\end{tabular}
}}
\label{tab:reranking-ml}
\vspace{-5pt}
\end{table*}

\begin{table*}[h]
\centering
\caption{The comparison of REKI and backbone reranking models on Amazon-Books dataset. Note that REKI-I and REKI-C both demonstrate significant improvements over the backbone rerankin models (t-test with $p$-value $<$ 0.05).}
\scalebox{0.88}{
\setlength{\tabcolsep}{1.3mm}{
\begin{tabular}{ccccccc|cccccc}
\toprule
\multirow{2}{*}{\textbf{\begin{tabular}[c]{@{}c@{}}Backbone\\ Model\end{tabular}}} & \multicolumn{3}{c}{\textbf{MAP@3}} & \multicolumn{3}{c|}{\textbf{MAP@7}} & \multicolumn{3}{c}
{\textbf{NDCG@3}} & \multicolumn{3}{c}{\textbf{NDCG@7}} \\
\cmidrule{2-13}
 & \textbf{base} & \textbf{REKI-C} & \textbf{REKI-I} & \textbf{base} & \textbf{REKI-C} & \textbf{REKI-I} & \textbf{base} & \textbf{REKI-C} & \textbf{REKI-I} & \textbf{base} & \textbf{REKI-C} & \textbf{REKI-I} \\
 \midrule
DLCM & 0.6365 & 0.6651 & \textbf{0.6654} & 0.6247 & 0.6488 & \textbf{0.6512} & 0.5755 & 0.6075 & \textbf{0.6109} & 0.6891 & 0.713 & \textbf{0.7142} \\
PRM & 0.6488 & 0.6738 & \textbf{0.6877} & 0.6359 & 0.6583 & \textbf{0.6722} & 0.5909 & 0.6191 & \textbf{0.6379} & 0.6983 & 0.7201 & \textbf{0.7312} \\
SetRank & 0.6509 & \textbf{0.6725} & 0.6711 & 0.6384 & \textbf{0.6573} & 0.6538 & 0.5947 & \textbf{0.6183} & 0.6137 & 0.7006 & \textbf{0.7191} & 0.7164 \\
MIR & 0.7178 & 0.7233 & \textbf{0.7241} & 0.7011 & 0.7069 & \textbf{0.7078} & 0.6747 & 0.6817 & \textbf{0.6837} & 0.7549 & 0.7591 & \textbf{0.7597} \\
\bottomrule
\end{tabular}
}}
\label{tab:reranking-ab}
\vspace{-5pt}
\end{table*}

\subsubsection{Improvement over Baselines (RQ2)} Next, we compare REKI with recent baselines using language models or sequence representation pretraining on the CTR task. The results are presented in Table~\ref{tab:baselines}. 
\begin{itemize}
    \item [(i)] We observe that REKI significantly outperforms models based on pretrained language models. For instance, with DIN as the backbone on Amazon-Books, REKI-I achieves a 0.91\% improvement in AUC and a 1.27\% improvement in LogLoss over the strongest baseline TALLRec. This validates the effectiveness of incorporation between LLMs and CRMs and the open-world knowledge extracted by REKI.
    \item [(ii)] When integrating PLMs into CRMs, the improvements brought by smaller PLMs are relatively modest. For instance, UniSRec, VQ-Rec, and P5, based on PLMs with less than one billion parameters, tend to exhibit poorer performance, even worse than the baseline DIN. This suggests that smaller language models may lack the capacity to effectively handle tasks related to user historical modeling or feature interaction in recommendation.
    \item [(iii)] Conversely, leveraging LLMs can lead to more substantial improvements, as seen in TALLRec and REKI. However, achieving effective enhancements with LLM embedding as initialization proves challenging, \eg, there is little difference between the results of LLM2DIN and DIN. This indicates that the parameters of large language models play a crucial role, and the information provided by mere initialization is relatively limited.
\end{itemize}


\begin{table}[h]
\centering
\caption{Comparison between REKI and baselines. }
\scalebox{1.0}{
\setlength{\tabcolsep}{1.2mm}{
\begin{tabular}{cccc|cc}
\toprule
\multirow{2}{*}{\textbf{Model}} & \multirow{2}{*}{\textbf{\begin{tabular}[c]{@{}c@{}}Backbone\\ PLM\end{tabular}}}& \multicolumn{2}{c|}{\textbf{MovieLens-1M}} & \multicolumn{2}{c}{\textbf{Amazon-Books}} \\
\cmidrule{3-6}
 & & \textbf{AUC} & \textbf{LogLoss} & \textbf{AUC} & \textbf{LogLoss} \\
 \midrule
UnisRec                         & BERT-110M                                                                         & 0.7702              & 0.5641              & 0.8196              & 0.5063              \\
VQ-Rec                          & BERT-110M                                                                         & 0.7707              & 0.5641              & 0.8226              & 0.5025              \\
P5                              & T5-223M                                                                           & 0.7790              & 0.5543              & 0.8333              & 0.4908              \\
LLM2DIN                         & ChatGLM-6B                                                                        & 0.7874              & 0.5473              & 0.8307              & 0.4930              \\
TALLRec                         & LLaMa2-7B                                                                         & \underline{0.7892}        & \underline{0.5451}        & \underline{0.8342}        & \underline{0.4862}        \\
base(DIN)                       & N/A                                                                               & 0.7863              & 0.5486              & 0.8304              & 0.4937              \\
REKI-C(DIN)                        & gpt-3.5-turbo                                                                        & \textbf{0.7952*}    & \textbf{0.5398*}    & \textbf{0.8396*}    & \textbf{0.4808*}   \\
REKI-I(DIN)                        & gpt-3.5-turbo                                                                        & \textbf{0.7961*}    & \textbf{0.5370*}    & \textbf{0.8418*}    & \textbf{0.4801*}   \\
\bottomrule
\end{tabular}
}}
\label{tab:baselines}
\footnotesize \flushleft\hspace{3cm} $*$ denotes statistically significant improvement over the second best baselines which is \\\flushleft\hspace{3cm}underlined (t-test with $p$-value $<$ 0.05).
\vspace{-5pt}
\end{table}


\subsubsection{Improvement over Other Knowledge-based Method (RQ3)} 
\begin{figure}[h]
    \centering
    \includegraphics[width=0.65\textwidth]{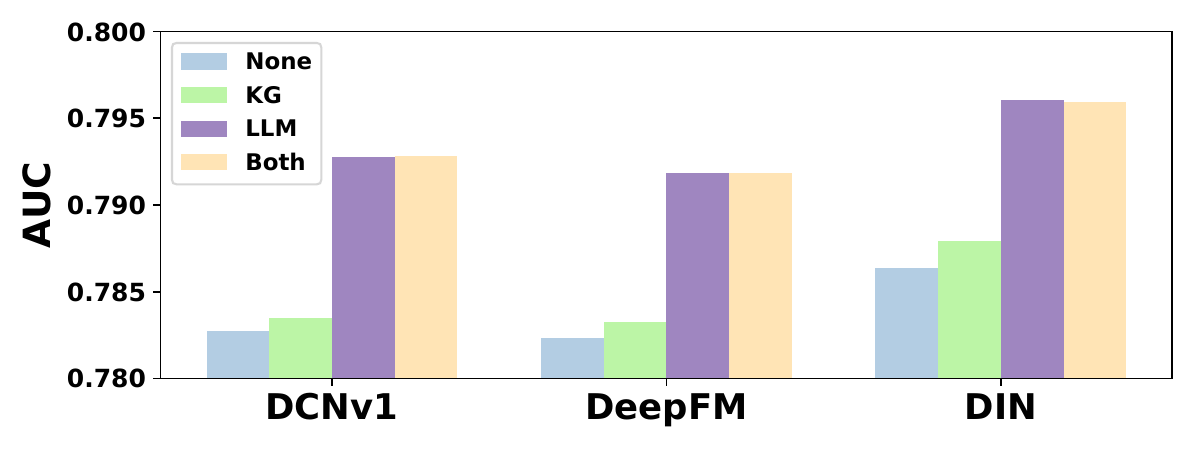}
    \caption{Comparison between knowledge from knowledge graph and LLM on MovieLens-1M dataset.}
\vspace{-5pt}
    \label{fig:kge}
\end{figure}

In Figure~\ref{fig:kge}, we compare knowledge from LLMs and other sources, such as knowledge graph (KG),  with DCNv1, DeepFM, and DIN as the backbone on MovieLens-1M dataset. We utilize a knowledge graph from LODrecsys~\cite{DOTD16}, which maps items of MovieLens-1M to DBPedia entities. Then, the entity embedding of each item is extracted following KTUP~\cite{cao2019unifying} and used as an additional feature for the backbone model. The legend ``\textbf{None}'' denotes the backbone model without knowledge enhancement, while ``\textbf{KG}'', ``\textbf{LLM}'', and ``\textbf{Both}'' represent the backbone model enhanced by knowledge from KG, LLM, and both sources, respectively. Note that the individual knowledge generated in \textbf{REKI-I} is utilized for \textbf{LLM}.

From Figure~\ref{fig:kge}, we notice that knowledge from both KG and LLM can bring performance improvements; however, the enhancement of KG is much smaller compared to that of LLM. This might be attributed to the fact that KG only possesses manually annotated item-side knowledge while it lacks knowledge for user preferences. According to our analysis in the next subsection~\ref{reas_fact_konwledge}, the knowledge for user preference usually contributes more gains compared to item factual knowledge. Furthermore, the simultaneous utilization of the two kinds of knowledge does not exhibit significant improvement over using LLM alone. This suggests that LLM may already encompass the typical knowledge within KG, making knowledge derived solely from LLM sufficient. 

\subsection{Deployment \& Online A/B Test (RQ4)}\label{sec:online}
\subsubsection{System Design}
\begin{figure}[h]
    \centering
    \includegraphics[width=0.8\textwidth]{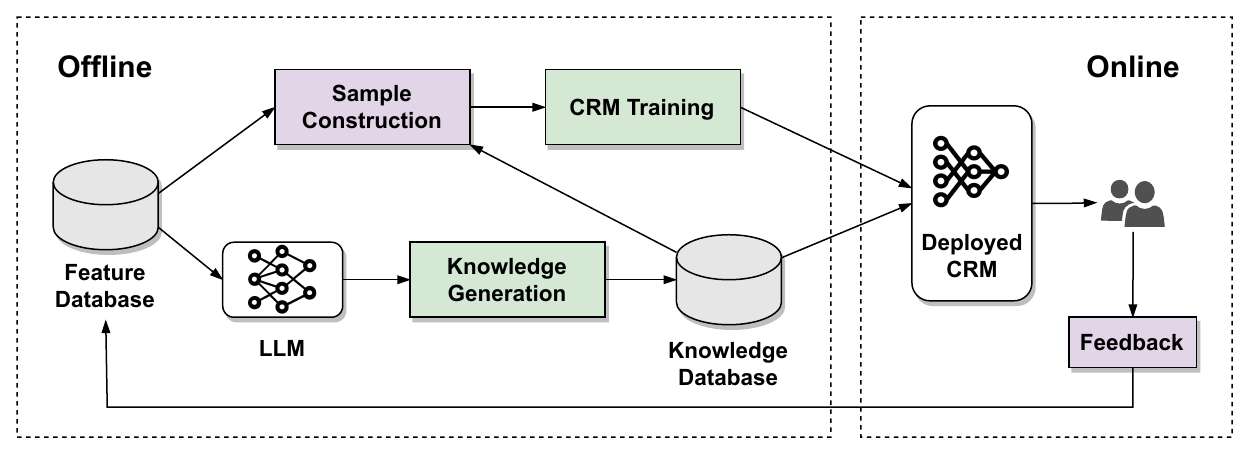}
    \caption{The system design for the deployment of REKI.}
\vspace{-5pt}
    \label{fig:system}
\end{figure}

In this section, we briefly introduce the system design of REKI during the deployment process, as illustrated in Figure~\ref{fig:system}. The entire system consists of two phases: the offline training phase and the online inference phase. The offline phase involves the knowledge generation of LLM and the training of the conventional recommendation model (CRM). During the knowledge generation, we extract features corresponding to users and items from the feature database based on the requirements of the scenario, which are used to construct prompts for LLM. Then, the LLM is employed to generate respective knowledge for users and items. Subsequently, a smaller language model encodes the generated knowledge into knowledge representations. These representations are stored in the knowledge database for subsequent training and inference of CRM. In the offline phase, we also conduct training for the CRM model. This process involves constructing training samples using the feature database and knowledge database and then completing the training. During the online inference phase, the CRM is deployed to the online system. The deployed CRM, along with data from the feature database and knowledge database, is utilized for inference, generating recommended results for users. User feedback is stored in the feature database for future training iterations.
\subsubsection{Online A/B Test}
To validate the effectiveness of REKI, we conducted two online experiments on Huawei's news and music platforms, respectively. In the news scenario with a relatively small number of users, we extracted knowledge for each user and item, deploying REKI-I. In the music scenario involving a larger number of users and items, we adopt REKI-C for efficient knowledge reasoning and generation. REKI shows significant improvement over the original baselines on the online A/B test for both scenarios. Now, REKI has already been deployed online and serves the main traffic.

In Huawei's news scenario, we focus on the task of news recall, for which we have adapted REKI-I into a recall model. During the knowledge reasoning and generation stage, we generate knowledge using Huawei's large language model PanGu\cite{zeng2021pangu}. Initially, we provide the LLM with the titles of the 20 most recently clicked news articles by a user. With the coarse-grained new categories, such as sports, finance, society, technology, etc., as scenario-specific factors, the LLM is required to infer the user's fine-grained preferences within those news categories. Once we obtain the user's fine-grained preferences, another smaller language model, Baize, designed by Huawei, encodes those preferences and candidate news articles. Those encodings are then used to calculate similarity scores, and news articles with higher similarity scores will be retrieved. The experimental group has deployed the aforementioned version of REKI-I, while the control group utilizes the original recall model as our baseline. During the online A/B test, REKI exhibited a 7\% improvement on the Recall metric compared with the baseline, resulting in significant business benefits. 

In the music scenario, we utilize REKI for the CTR prediction task, which we have mentioned in our experiments. This scenario involves a larger number of users and items, leading us to adopt REKI-C. In this scenario, 30\% of users were randomly selected into the experimental group, and another 30\% were in the control group. Both groups used the same base model for generating recommendations. The difference lies in their inputs, where the input of the experimental group included the knowledge representation augmented by REKI-C. In large-scale domains, storage efficiency is also a crucial aspect that we need to consider. Therefore, we employ Principal Component Analysis (PCA) to reduce the dimension of knowledge representations to 64 and store them in a vector database. Additionally, we adopt incremental update, which only processes the updated items and users daily and adds them to the vector database. In a 7-day online A/B test, REKI demonstrated a 1.99\% increase in song play count, a 1.73\% increase in the number of devices for song playback, and a 2.04\% increase in total duration. Now, REKI has already been deployed online and serves the main traffic in this scenario. This indicates that REKI can be successfully implemented in industrial settings and improve the recommendation experience for real-world users. 


\subsection{Efficiency Study (RQ5)}\label{sec:efficiency}
\subsubsection{Offline Efficiency}\label{sec:infer_compar}
To study the offline efficiency of our proposed model, we first compared the performance and efficiency of two variants of REKI, namely \textbf{REKI-I} and \textbf{REKI-C}, which utilize individual and collective forms of knowledge, respectively, with base model DIN and the strongest baseline TALLREC. REKI-I requires generating individual knowledge using LLM for each item and user, while REKI-C generates collective knowledge for clusters of items and users, reducing the number of knowledge generations. For both REKI-I and REKI-C, we employ three different sizes of encoders: ChatGLM-6B~\cite{du2022glm}, BERT-110M~\cite{bert}, and tinyBERT-14.7M~\cite{jiao2019tinybert}. Regarding offline efficiency, we compared the number of knowledge generation calls for the entire dataset (\textbf{\# call} in the table) and the training time per batch for each model, which can be regarded as measures of offline resource consumption (both time and computational resource). For REKI-I and REKI-C, \textbf{\# call} represents the number of knowledge generation calls on LLM API. For baseline TALLREC, \textbf{\# call} denotes the number of times the LLM is invoked to generate the final CTR predictions, corresponding to the sum of training and testing data samples. All experiments in Table~\ref{tab:inference_comp} are conducted on the same device and environment, and the batch size of training is set to 256. The only exception is that the GPU memory size limitation prevents us from setting the batch size of TALLREC to 256, so we present the training time with the batch size set to 1. From these experiments, we draw the following conclusions.

\begin{table*}[h]
\centering
\caption{Offline efficiency comparison amongst variants of REKI and baselines.}
\scalebox{0.9}{
\setlength{\tabcolsep}{1.3mm}{
\begin{tabular}{cccccc|cccc}
\toprule
\multirow{2}{*}{\textbf{Model}} & \multirow{2}{*}{\textbf{Encoder}} & \multicolumn{4}{c|}{\textbf{MovieLens-1M}} & \multicolumn{4}{c}{\textbf{Amazon-books}} \\
\cmidrule{3-10}
 &  & \textbf{\# call} & \textbf{training time (s)} & \textbf{AUC} & \textbf{Logloss} & \textbf{\# call} & \textbf{training time (s)} & \textbf{AUC} & \textbf{Logloss} \\
 \midrule
\textbf{DIN} & / & / & $8.41\times10^{-3}$ & 0.7863 & 0.5486 & / & $1.46\times10^{-2}$ & 0.8304 & 0.4937 \\
\textbf{TALLREC} & / & 167167 & $9.76\times10^{-1}$ & 0.7892 & 0.5451 & 268029 & $8.91\times10^{-1}$ & 0.8342 & 0.4862 \\
\midrule
\multirow{3}{*}{\textbf{REKI-I}} & \textbf{ChatGLM-6B} & 9923 & $9.59\times10^{-2}$ & 0.7961 & 0.5370 & 29238 & $9.68\times10^{-2}$ & 0.8418 & 0.4801 \\

 & BERT-110M & 9923 & $2.45\times10^{-2}$ & 0.7947 & 0.5412 & 29238 & $2.42\times10^{-2}$ & 0.8381 & 0.4839 \\
 & tinyBERT-14.5M & 9923 & $1.48\times10^{-2}$ & 0.7907 & 0.5434 & 29238 & $1.47\times10^{-2}$ & 0.8358 & 0.4877 \\
 \bottomrule
\multirow{3}{*}{\textbf{REKI-C}} & ChatGLM-6B & 377 & $2.19\times10^{-1}$ & 0.7959 & 0.5387 & 2080 & $2.22\times10^{-1}$ & 0.8388 & 0.4814 \\
 & \textbf{BERT-110M} & 377 & $5.08\times10^{-2}$ & 0.7952 & 0.5398 & 2080 & $5.32\times10^{-2}$ & 0.8396 & 0.4808 \\
 & tinyBERT-14.5M & 377 & $2.77\times10^{-2}$ & 0.7928 & 0.5428 & 2080 & $2.75\times10^{-2}$ & 0.8378 & 0.4848\\
 \bottomrule
\end{tabular}
}}
\label{tab:inference_comp}
\end{table*}

Firstly, REKI demonstrates significantly higher efficiency in both time and computation resource consumption compared to approaches like TALLREC that directly utilize LLMs for recommendation. This efficiency arises from the fact that REKI-I only needs to generate knowledge for each user and item, whereas TALLREC needs to generate knowledge for all samples, and the quantity of samples is typically much larger than that of users and items. For instance, in the \textbf{\# call} metric of MovieLens-1M dataset, the knowledge generation frequency of REKI-I is one-sixteenth of TALLREC, which indicates that REKI-I incurs significantly fewer computational resources and less time consumption than TALLREC. Moreover, regarding both training times, REKI-I also outperforms TALLREC by a significant margin, roughly being only one-tenth of TALLREC. This also provides strong evidence of REKI's efficiency over TALLREC.

Secondly, for knowledge generation, collective knowledge extraction used by REKI-C effectively reduces the number of knowledge generation calls without losing much accuracy. For example, on MovieLens, \textbf{\# call} for REKI-C is one twenty-sixth of REKI-I, but the two variants show comparable prediction accuracy. Moreover, with more miniature encoders such as BERT and tinyBERT, REKI-C even outperforms REKI-I. Considering that knowledge generation involves larger language models, which are slower, using REKI-C can effectively improve the deployment efficiency of the model.

Lastly, with the same encoder, REKI-C exhibits longer training times compared to REKI-I, but REKI-C is more friendly to more miniature encoders. With the same encoder, REKI-C has approximately twice the training time of REKI-I, as REKI-C generates and utilizes more additional features. However, REKI-C is more friendly to smaller encoders. With encoders like BERT and tinyBERT, REKI-C exhibits an improvement over REKI-I, possibly because collective knowledge is more accessible for smaller encoders to comprehend. Therefore, in practical use, adopting REKI-C with a smaller encoder model, such as BERT, can lead to significant accelerations in knowledge generation and training time compared to REKI-I with ChatGLM, thereby enhancing model efficiency. This is why we employ ChatGLM-6B and BERT as default encoders for REKI-I and REKI-C in the preceding experiments.

\subsubsection{Inference Latency}\label{sec:speed_up_efficiency}
To quantify the actual inference latency of REKI and assess the effectiveness of the acceleration strategies, we compare the inference time of REKI-I based on DIN with LLM API, strongest baseline TALLRec, and base DIN model in Table~\ref{tab:inference}. For \textbf{LLM API}, we follow the zero-shot user rating prediction in~\cite{kang2023llms} that provides the user viewing history and ratings as prompt and invokes the API to predict user ratings on candidate items. Since this approach does not allow setting a batch size, the table presents the average response time per sample. For REKI, we evaluate the two acceleration strategies as introduced in Section~\ref{sec:speed_up}: $\textbf{REKI}_{w/\,HEIN}$, where the hybridized expert integration network (HEIN) participates in the inference stage, and $\textbf{REKI}_{w/o\,HEIN}$, where HEIN is detached from inference. As we have different version of knowledge generation, which also impact the inference speed, we present four variants of REKI in Table~\ref{tab:inference}: $\textbf{REKI-I}_{w/\,HEIN}$, $\textbf{REKI-I}_{w/o\,HEIN}$, $\textbf{REKI-C}_{w/\,HEIN}$, and $\textbf{REKI-C}_{w/o\,HEIN}$. The experiments of REKI and base model are all conducted on the same device and environment, with a batch size of 256. With the same device, we also test \textbf{TALLRec} and only showcase the average inference time per sample, since 32G memory cannot handle LLM with a batch size of 256. Table~\ref{tab:inference} presents the average inference time, from which we draw the following conclusions.

\begin{table}[h]
\centering
\caption{The comparison of inference time (s).}
\scalebox{1.0}{
\setlength{\tabcolsep}{1.0mm}{
\begin{tabular}{ccc}
\toprule
\textbf{Model}      & \textbf{MovieLens-1M} & \textbf{Amazon-Books} \\
\midrule
LLM API   & $5.54$ & $4.11$ \\
TALLRec   &  $7.63\times10^{-1}$ & $7.97\times10^{-1}$ \\
$\text{REKI-I}_{w/\,\,HEIN}$  & $8.08\times10^{-2}$ & $9.39\times10^{-2}$ \\

$\text{REKI-C}_{w/\,\,HEIN}$  & $4.63\times10^{-2}$ & $4.70\times10^{-2}$ \\
$\text{REKI-I}_{w/o\,\,HEIN}$ & $6.64\times10^{-3}$ & $1.11\times10^{-2}$\\

$\text{REKI-C}_{w/o\,\,HEIN}$ & $6.56\times10^{-3}$ & $1.11\times10^{-2}$\\
base DIN     & $6.42\times10^{-3}$ & $1.09\times10^{-2}$\\
\bottomrule
\end{tabular}
}}
\vspace{-5pt}
\label{tab:inference}
\end{table}

Firstly, adopting LLM for direct inference is not feasible for RSs due to its large computational latency. The response latency of LLM API is 4-6 seconds, which does not meet the real-time requirement of RSs, which typically demands a response latency of within 100ms. Even if we finetune a relatively small LLM, such as TALLRec based on LLaMa2-7B, its inference latency is still close to 1 second, which is unbearable for industrial scenarios. 

Secondly, both acceleration methods of REKI achieve an inference time within 100ms, satisfying the low latency requirement. When using the prestoring knowledge representation approach, $\textbf{REKI-C}_{w/\,HEIN}$ has a shorter inference time than $\textbf{REKI-I}_{w/o\,HEIN}$. This is because $\textbf{REKI-C}_{w/o\,HEIN}$ employs BERT as the encoder, resulting in knowledge representation with a dimension of 768. On the other hand, $\textbf{REKI-I}_{w/o\,HEIN}$ uses ChatGLM-6B as the encoder, producing representations with a dimension of 4096, which requires more processing time. Importantly, if we employ the approach of prestoring user and item augmented vectors, the knowledge generation method no longer has a significant impact. In other words, the actual inference time of $\text{REKI-I}_{w/o\,HEIN}$ and  $\text{REKI-C}_{w/o\,HEIN}$ become nearly identical, almost the same as that of the backbone model. This demonstrates the effectiveness of our proposed REKI and acceleration strategies.

\subsection{Ablation Study (RQ6)}
\subsubsection{User and Item Knowledge}\label{reas_fact_konwledge}

\begin{figure}[h]
    \centering
    \includegraphics[width=0.65\textwidth]{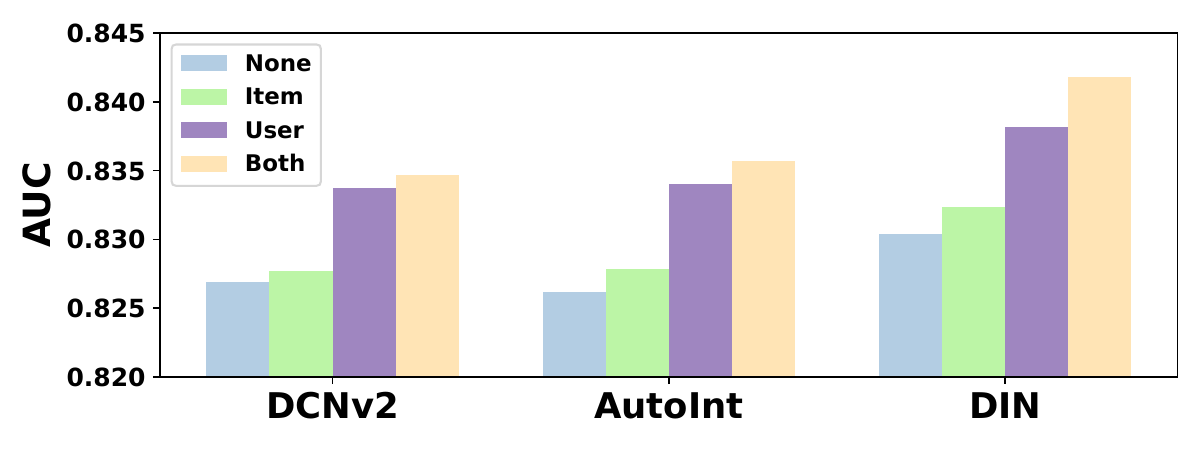}
    \caption{Ablation study about user and item knowledge on Amazon-Books dataset.}
\vspace{-5pt}
    \label{fig:knowledge_abaltion}
\end{figure}

To study the impact of knowledge generated by LLMs, we conduct an ablation study on user and item knowledge on the Amazon-Books dataset. 
The two knowledge generation approaches, individual knowledge extraction (REKI-I) and collective knowledge extraction (REKI-C), both produce user and item knowledge, so here we use REKI-I as an example.
We select DCNv2, AutoInt, and DIN as backbone models and compare their performance with different knowledge enhancements, as shown in Figure~\ref{fig:knowledge_abaltion}. The legend ``\textbf{None}'' represents the backbone model without any knowledge enhancement. ``\textbf{Item}'' and ``\textbf{User}'' indicate the backbone models enhanced with item knowledge and user knowledge, respectively, while ``\textbf{Both}'' represents the joint use of both knowledge types.

From Figure~\ref{fig:knowledge_abaltion}, we observe that both item knowledge and user knowledge can improve the performance of backbone models, with user knowledge exhibiting a larger improvement. This could be attributed to the fact that user knowledge inferred by the LLMs captures in-depth user preferences, thus compensating for the backbone model's limitations in reasoning underlying motives and intentions. Additionally, the joint use of both user and item enhancements outperforms using either one alone, even achieving a synergistic effect where $1+1>2$. One possible explanation is that user knowledge contains external information that is not explicitly present in the raw data. When used independently, this external knowledge could not be matched with candidate items. However, combining the externally generated item factual knowledge from the LLMs aligned with user knowledge allows RSs to gain a better understanding of items according to user preferences.

\begin{table}[h]
\centering
\caption{Impact of different knowledge encoders. 
}
\scalebox{1.0}{
\setlength{\tabcolsep}{0.9mm}{
\begin{tabular}{ccccc|cccc}
\toprule
\multirow{3}{*}{\textbf{Variants}} & \multicolumn{4}{c|}{\textbf{MoiveLens-1M}} & \multicolumn{4}{c}{\textbf{Amazon-Books}} \\
\cmidrule{2-9}
 & \multicolumn{2}{c}{\textbf{BERT}} & \multicolumn{2}{c|}{\textbf{ChatGLM}} & \multicolumn{2}{c}{\textbf{BERT}} & \multicolumn{2}{c}{\textbf{ChatGLM}} \\
 \cmidrule{2-9}
 & \textbf{AUC} & \textbf{LogLoss} & \textbf{AUC} & \textbf{LogLoss} & \textbf{AUC} & \textbf{LogLoss} & \textbf{AUC} & \textbf{LogLoss} \\
 \midrule
base(DIN) & 0.7863 & 0.5486 & 0.7863 & 0.5486 & 0.8304 & 0.4937 & 0.8304 & 0.4937 \\
REKI(LR) & 0.7589 & 0.5811 & 0.7720 & 0.5674 & 0.7375 & 0.5861 & 0.7560 & 0.5718 \\
REKI(MLP) & 0.7754 & 0.5588 & 0.7816 & 0.5559 & 0.7424 & 0.5834 & 0.7571 & 0.5763 \\
REKI-MLP & 0.7934 & 0.5411 & 0.7939 & 0.5401 & 0.8357 & 0.4880 & 0.8370 & 0.4843 \\
REKI-MoE & {\underline{0.7946}} & \textbf{0.5391} & {\underline{0.7950}} & {\underline{0.5394}} & {\underline{0.8371}} & \textbf{0.4841} & {\underline{0.8388}} & {\underline{0.4823}} \\
REKI & \textbf{0.7947} & {\underline{0.5402}} & \textbf{0.7961} & \textbf{0.5370} & \textbf{0.8374} & {\underline{0.4843}} & \textbf{0.8418} & \textbf{0.4801}
\\
\bottomrule
\end{tabular}
}}
\label{tab:text_encoder}
\end{table}

\subsubsection{Knowledge Encoders and Semantic Transformation}\label{sec:knowledge_encoder}
We employ two different language models, BERT~\cite{bert} and ChatGLM~\cite{du2022glm}, to investigate the impact of different knowledge encoders on \textbf{REKI-I}'s performance. Additionally, we design several variants to demonstrate how the representations generated by knowledge encoders are utilized. \textbf{REKI(LR)} applies average pooling to token representations from the knowledge encoder and directly feeds the result into a linear layer to obtain prediction scores, without utilizing a backbone CTR model. \textbf{REKI(MLP)} replaces the linear layer of REKI(LR) with an MLP. \textbf{REKI-MLP} and \textbf{REKI-MoE} replace the HEIN module with an MLP and a Mixture-of-Experts (MoE), respectively. The original REKI and the two variants, REKI-MLP and REKI-MoE, all adopt DIN as the backbone model. Table~\ref{tab:text_encoder} presents their performance, from which we draw the following conclusions. 

Firstly, we can observe that, overall, variants with ChatGLM as the knowledge encoder outperform those with BERT. The performance of REKI(LR) and REKI(MLP) can be considered as a measure of the quality of the encoded representations, since they directly adopt the representations for prediction. Considering REKI(LR) and REKI(MLP), 
the superior performance of ChatGLM over BERT indicates that ChatGLM performs better in preserving the information within knowledge from LLMs, which may be attributed to the larger size and better text comprehension of ChatGLM (6 billion) compared to BERT (110 million). With ChatGLM on MovieLens-1M, REKI(MLP) is even close to base(DIN), validating the effectiveness of our generated open-world knowledge.



Secondly, the performance of REKI benefits from complex semantic transformation structures, but the knowledge encoder also limits it. The results on ChatGLM show that a simple MLP is less effective than MoE and our designed HEIN outperforms the MoE, indicating that the transformation from semantic space to recommendation space entails a complex network structure. However, with BERT as knowledge encoder, REKI-MoE and REKI-MLP exhibit similar performance, suggesting that information from BERT is limited and using an MoE is sufficient in this case.

\section{Broader Impact}
When incorporating LLMs into RSs, it is imperative to give serious consideration to privacy and security issues. While LLMs offer remarkable capabilities, they also come with potential risks of compromising user privacy and generating harmful content~\cite{pan2020privacy,Brown2022privacy,weidinger2021ethical}. We also consider these two issues while designing our proposed framework, REKI.

Regarding privacy, REKI leverages LLMs' reasoning ability and factual knowledge without finetuning LLMs, ensuring that LLMs do not retain or remember user-specific data. Besides, while experiments may involve the use of external APIs on public datasets, real-world implementations usually rely on in-house models, \eg, we utilize Huawei's own LLM PanGu\cite{zeng2021pangu} for online A/B test as described in section~\ref{sec:online}, preventing the leakage of user privacy information through external APIs.

Furthermore, compared to methods that directly display LLM-generated content to users~\cite{chatrec,liu2023chatgpt,dai2023uncovering}, REKI takes a  more proactive stance to mitigate concerns about harmful content and hallucination knowledge generated by LLMs. REKI first converts the textual content from LLMs to robust representations and then incorporates those representations into traditional RSs. This integration avoids displaying
harmful or misleading content generated by LLMs while enabling us to utilize filtering mechanisms commonly used in traditional RSs to screen out potential harmful item recommendations.

Through those measures, REKI strives to deliver robust recommendation performance while safeguarding user data privacy and the security of recommendation content. 


\section{Conclusion}

Our work presents REKI, a framework for effectively incorporating open-world knowledge into recommender systems by exploiting large language models. In REKI, we introduce factorization prompting for accurate knowledge reasoning on users and items and develop individual and collective knowledge extraction for different scales of scenarios, effectively reducing offline resource consumption. Subsequently, a hybridized expert-integrated network is designed to efficiently transform the generated user and item knowledge into augmented vectors to enhance any conventional recommendation model. We also ensure efficient inference by preprocessing and prestoring the generated knowledge. Extensive experiments demonstrate that REKI significantly outperforms the state-of-the-art baselines and is compatible with various recommendation algorithms and tasks. REKI has been deployed to Huawei's news and music recommendation platforms, gaining a 7\% and 1.99\% improvement during the online A/B test. In this work, we have explored how to reduce the knowledge LLMs need to generate to improve efficiency and reduce resource consumption. In the future, we will investigate ways to accelerate the recommendation knowledge generation process of LLMs, such as through quantization and speculative decoding, to further reduce the resource consumption of deploying LLMs in recommendation systems.

%


\bibliographystyle{ACM-Reference-Format}
\bibliography{myref}


\begin{thebibliography}{96}


\ifx \showCODEN    \undefined \def \showCODEN     #1{\unskip}     \fi
\ifx \showDOI      \undefined \def \showDOI       #1{#1}\fi
\ifx \showISBNx    \undefined \def \showISBNx     #1{\unskip}     \fi
\ifx \showISBNxiii \undefined \def \showISBNxiii  #1{\unskip}     \fi
\ifx \showISSN     \undefined \def \showISSN      #1{\unskip}     \fi
\ifx \showLCCN     \undefined \def \showLCCN      #1{\unskip}     \fi
\ifx \shownote     \undefined \def \shownote      #1{#1}          \fi
\ifx \showarticletitle \undefined \def \showarticletitle #1{#1}   \fi
\ifx \showURL      \undefined \def \showURL       {\relax}        \fi
\providecommand\bibfield[2]{#2}
\providecommand\bibinfo[2]{#2}
\providecommand\natexlab[1]{#1}
\providecommand\showeprint[2][]{arXiv:#2}

\bibitem[Ahmed et~al\mbox{.}(2020)]%
        {ahmed2020k}
\bibfield{author}{\bibinfo{person}{Mohiuddin Ahmed}, \bibinfo{person}{Raihan Seraj}, {and} \bibinfo{person}{Syed Mohammed~Shamsul Islam}.} \bibinfo{year}{2020}\natexlab{}.
\newblock \showarticletitle{The k-means algorithm: A comprehensive survey and performance evaluation}.
\newblock \bibinfo{journal}{\emph{Electronics}} \bibinfo{volume}{9}, \bibinfo{number}{8} (\bibinfo{year}{2020}), \bibinfo{pages}{1295}.
\newblock


\bibitem[Ai et~al\mbox{.}(2018)]%
        {ai2018learning}
\bibfield{author}{\bibinfo{person}{Qingyao Ai}, \bibinfo{person}{Keping Bi}, \bibinfo{person}{Jiafeng Guo}, {and} \bibinfo{person}{W~Bruce Croft}.} \bibinfo{year}{2018}\natexlab{}.
\newblock \showarticletitle{Learning a deep listwise context model for ranking refinement}. In \bibinfo{booktitle}{\emph{The 41st international ACM SIGIR conference on research \& development in information retrieval}}. \bibinfo{pages}{135--144}.
\newblock


\bibitem[Bao et~al\mbox{.}(2023)]%
        {bao2023tallrec}
\bibfield{author}{\bibinfo{person}{Keqin Bao}, \bibinfo{person}{Jizhi Zhang}, \bibinfo{person}{Yang Zhang}, \bibinfo{person}{Wenjie Wang}, \bibinfo{person}{Fuli Feng}, {and} \bibinfo{person}{Xiangnan He}.} \bibinfo{year}{2023}\natexlab{}.
\newblock \showarticletitle{Tallrec: An effective and efficient tuning framework to align large language model with recommendation}.
\newblock \bibinfo{journal}{\emph{arXiv preprint arXiv:2305.00447}} (\bibinfo{year}{2023}).
\newblock


\bibitem[Brown et~al\mbox{.}(2022)]%
        {Brown2022privacy}
\bibfield{author}{\bibinfo{person}{Hannah Brown}, \bibinfo{person}{Katherine Lee}, \bibinfo{person}{Fatemehsadat Mireshghallah}, \bibinfo{person}{Reza Shokri}, {and} \bibinfo{person}{Florian Tram\`{e}r}.} \bibinfo{year}{2022}\natexlab{}.
\newblock \showarticletitle{What Does It Mean for a Language Model to Preserve Privacy?}. In \bibinfo{booktitle}{\emph{Proceedings of the 2022 ACM Conference on Fairness, Accountability, and Transparency}} \emph{(\bibinfo{series}{FAccT '22})}. \bibinfo{pages}{2280–2292}.
\newblock


\bibitem[Bubeck et~al\mbox{.}(2023)]%
        {bubeck2023sparks}
\bibfield{author}{\bibinfo{person}{S{\'e}bastien Bubeck}, \bibinfo{person}{Varun Chandrasekaran}, \bibinfo{person}{Ronen Eldan}, \bibinfo{person}{Johannes Gehrke}, \bibinfo{person}{Eric Horvitz}, \bibinfo{person}{Ece Kamar}, \bibinfo{person}{Peter Lee}, \bibinfo{person}{Yin~Tat Lee}, \bibinfo{person}{Yuanzhi Li}, \bibinfo{person}{Scott Lundberg}, {et~al\mbox{.}}} \bibinfo{year}{2023}\natexlab{}.
\newblock \showarticletitle{Sparks of artificial general intelligence: Early experiments with gpt-4}.
\newblock \bibinfo{journal}{\emph{arXiv preprint arXiv:2303.12712}} (\bibinfo{year}{2023}).
\newblock


\bibitem[Cao et~al\mbox{.}(2019)]%
        {cao2019unifying}
\bibfield{author}{\bibinfo{person}{Yixin Cao}, \bibinfo{person}{Xiang Wang}, \bibinfo{person}{Xiangnan He}, \bibinfo{person}{Zikun Hu}, {and} \bibinfo{person}{Tat-Seng Chua}.} \bibinfo{year}{2019}\natexlab{}.
\newblock \showarticletitle{Unifying knowledge graph learning and recommendation: Towards a better understanding of user preferences}. In \bibinfo{booktitle}{\emph{The world wide web conference}}. \bibinfo{pages}{151--161}.
\newblock


\bibitem[Chen et~al\mbox{.}(2023)]%
        {chen2023large}
\bibfield{author}{\bibinfo{person}{Jin Chen}, \bibinfo{person}{Zheng Liu}, \bibinfo{person}{Xu Huang}, \bibinfo{person}{Chenwang Wu}, \bibinfo{person}{Qi Liu}, \bibinfo{person}{Gangwei Jiang}, \bibinfo{person}{Yuanhao Pu}, \bibinfo{person}{Yuxuan Lei}, \bibinfo{person}{Xiaolong Chen}, \bibinfo{person}{Xingmei Wang}, {et~al\mbox{.}}} \bibinfo{year}{2023}\natexlab{}.
\newblock \showarticletitle{When large language models meet personalization: Perspectives of challenges and opportunities}.
\newblock \bibinfo{journal}{\emph{arXiv preprint arXiv:2307.16376}} (\bibinfo{year}{2023}).
\newblock


\bibitem[Cui et~al\mbox{.}(2022)]%
        {m6rec}
\bibfield{author}{\bibinfo{person}{Zeyu Cui}, \bibinfo{person}{Jianxin Ma}, \bibinfo{person}{Chang Zhou}, \bibinfo{person}{Jingren Zhou}, {and} \bibinfo{person}{Hongxia Yang}.} \bibinfo{year}{2022}\natexlab{}.
\newblock \showarticletitle{M6-Rec: Generative Pretrained Language Models are Open-Ended Recommender Systems}.
\newblock \bibinfo{journal}{\emph{arXiv preprint arXiv:2205.08084}} (\bibinfo{year}{2022}).
\newblock


\bibitem[Dai et~al\mbox{.}(2023)]%
        {dai2023uncovering}
\bibfield{author}{\bibinfo{person}{Sunhao Dai}, \bibinfo{person}{Ninglu Shao}, \bibinfo{person}{Haiyuan Zhao}, \bibinfo{person}{Weijie Yu}, \bibinfo{person}{Zihua Si}, \bibinfo{person}{Chen Xu}, \bibinfo{person}{Zhongxiang Sun}, \bibinfo{person}{Xiao Zhang}, {and} \bibinfo{person}{Jun Xu}.} \bibinfo{year}{2023}\natexlab{}.
\newblock \showarticletitle{Uncovering ChatGPT's Capabilities in Recommender Systems}.
\newblock \bibinfo{journal}{\emph{arXiv preprint arXiv:2305.02182}} (\bibinfo{year}{2023}).
\newblock


\bibitem[{Di Noia} et~al\mbox{.}(2016)]%
        {DOTD16}
\bibfield{author}{\bibinfo{person}{Tommaso {Di Noia}}, \bibinfo{person}{Vito~Claudio Ostuni}, \bibinfo{person}{Paolo Tomeo}, {and} \bibinfo{person}{Eugenio {Di Sciascio}}.} \bibinfo{year}{2016}\natexlab{}.
\newblock \showarticletitle{SPRank: Semantic Path-based Ranking for Top-N Recommendations using Linked Open Data}.
\newblock \bibinfo{journal}{\emph{ACM Transactions on Intelligent Systems and Technology (TIST)}} (\bibinfo{year}{2016}).
\newblock


\bibitem[Ding et~al\mbox{.}(2021)]%
        {zesrec}
\bibfield{author}{\bibinfo{person}{Hao Ding}, \bibinfo{person}{Yifei Ma}, \bibinfo{person}{Anoop Deoras}, \bibinfo{person}{Yuyang Wang}, {and} \bibinfo{person}{Hao Wang}.} \bibinfo{year}{2021}\natexlab{}.
\newblock \showarticletitle{Zero-shot recommender systems}.
\newblock \bibinfo{journal}{\emph{arXiv preprint arXiv:2105.08318}} (\bibinfo{year}{2021}).
\newblock


\bibitem[Du et~al\mbox{.}(2022)]%
        {du2022glm}
\bibfield{author}{\bibinfo{person}{Zhengxiao Du}, \bibinfo{person}{Yujie Qian}, \bibinfo{person}{Xiao Liu}, \bibinfo{person}{Ming Ding}, \bibinfo{person}{Jiezhong Qiu}, \bibinfo{person}{Zhilin Yang}, {and} \bibinfo{person}{Jie Tang}.} \bibinfo{year}{2022}\natexlab{}.
\newblock \showarticletitle{GLM: General Language Model Pretraining with Autoregressive Blank Infilling}. In \bibinfo{booktitle}{\emph{Proceedings of the 60th Annual Meeting of the Association for Computational Linguistics (Volume 1: Long Papers)}}. \bibinfo{pages}{320--335}.
\newblock


\bibitem[Fan et~al\mbox{.}(2023)]%
        {fan2023recommender}
\bibfield{author}{\bibinfo{person}{Wenqi Fan}, \bibinfo{person}{Zihuai Zhao}, \bibinfo{person}{Jiatong Li}, \bibinfo{person}{Yunqing Liu}, \bibinfo{person}{Xiaowei Mei}, \bibinfo{person}{Yiqi Wang}, \bibinfo{person}{Jiliang Tang}, {and} \bibinfo{person}{Qing Li}.} \bibinfo{year}{2023}\natexlab{}.
\newblock \showarticletitle{Recommender systems in the era of large language models (llms)}.
\newblock \bibinfo{journal}{\emph{arXiv preprint arXiv:2307.02046}} (\bibinfo{year}{2023}).
\newblock


\bibitem[Friedman et~al\mbox{.}(2023)]%
        {friedman2023leveraging}
\bibfield{author}{\bibinfo{person}{Luke Friedman}, \bibinfo{person}{Sameer Ahuja}, \bibinfo{person}{David Allen}, \bibinfo{person}{Terry Tan}, \bibinfo{person}{Hakim Sidahmed}, \bibinfo{person}{Changbo Long}, \bibinfo{person}{Jun Xie}, \bibinfo{person}{Gabriel Schubiner}, \bibinfo{person}{Ajay Patel}, \bibinfo{person}{Harsh Lara}, {et~al\mbox{.}}} \bibinfo{year}{2023}\natexlab{}.
\newblock \showarticletitle{Leveraging Large Language Models in Conversational Recommender Systems}.
\newblock \bibinfo{journal}{\emph{arXiv preprint arXiv:2305.07961}} (\bibinfo{year}{2023}).
\newblock


\bibitem[Gao et~al\mbox{.}(2023)]%
        {chatrec}
\bibfield{author}{\bibinfo{person}{Yunfan Gao}, \bibinfo{person}{Tao Sheng}, \bibinfo{person}{Youlin Xiang}, \bibinfo{person}{Yun Xiong}, \bibinfo{person}{Haofen Wang}, {and} \bibinfo{person}{Jiawei Zhang}.} \bibinfo{year}{2023}\natexlab{}.
\newblock \showarticletitle{Chat-REC: Towards Interactive and Explainable LLMs-Augmented Recommender System}.
\newblock \bibinfo{journal}{\emph{arXiv preprint arXiv:2303.14524}} (\bibinfo{year}{2023}).
\newblock


\bibitem[Geng et~al\mbox{.}(2022)]%
        {p5}
\bibfield{author}{\bibinfo{person}{Shijie Geng}, \bibinfo{person}{Shuchang Liu}, \bibinfo{person}{Zuohui Fu}, \bibinfo{person}{Yingqiang Ge}, {and} \bibinfo{person}{Yongfeng Zhang}.} \bibinfo{year}{2022}\natexlab{}.
\newblock \showarticletitle{Recommendation as Language Processing (RLP): A Unified Pretrain, Personalized Prompt and Predict Paradigm (P5)}. In \bibinfo{booktitle}{\emph{Proceedings of the 16th ACM Conference on Recommender Systems}}. \bibinfo{pages}{299–315}.
\newblock


\bibitem[Gong et~al\mbox{.}(2023)]%
        {gong2023unified}
\bibfield{author}{\bibinfo{person}{Yuqi Gong}, \bibinfo{person}{Xichen Ding}, \bibinfo{person}{Yehui Su}, \bibinfo{person}{Kaiming Shen}, \bibinfo{person}{Zhongyi Liu}, {and} \bibinfo{person}{Guannan Zhang}.} \bibinfo{year}{2023}\natexlab{}.
\newblock \showarticletitle{An Unified Search and Recommendation Foundation Model for Cold-Start Scenario}. In \bibinfo{booktitle}{\emph{Proceedings of the 32nd ACM International Conference on Information and Knowledge Management}} \emph{(\bibinfo{series}{CIKM '23})}. \bibinfo{pages}{4595–4601}.
\newblock


\bibitem[Guo et~al\mbox{.}(2017)]%
        {DeepFM}
\bibfield{author}{\bibinfo{person}{Huifeng Guo}, \bibinfo{person}{Ruiming Tang}, \bibinfo{person}{Yunming Ye}, \bibinfo{person}{Zhenguo Li}, {and} \bibinfo{person}{Xiuqiang He}.} \bibinfo{year}{2017}\natexlab{}.
\newblock \showarticletitle{DeepFM: A Factorization-Machine Based Neural Network for CTR Prediction}. In \bibinfo{booktitle}{\emph{Proceedings of the 26th International Joint Conference on Artificial Intelligence}} (Melbourne, Australia). \bibinfo{pages}{1725–1731}.
\newblock


\bibitem[Guo et~al\mbox{.}(2020)]%
        {guo2020survey}
\bibfield{author}{\bibinfo{person}{Qingyu Guo}, \bibinfo{person}{Fuzhen Zhuang}, \bibinfo{person}{Chuan Qin}, \bibinfo{person}{Hengshu Zhu}, \bibinfo{person}{Xing Xie}, \bibinfo{person}{Hui Xiong}, {and} \bibinfo{person}{Qing He}.} \bibinfo{year}{2020}\natexlab{}.
\newblock \showarticletitle{A survey on knowledge graph-based recommender systems}.
\newblock \bibinfo{journal}{\emph{IEEE Transactions on Knowledge and Data Engineering}} \bibinfo{volume}{34}, \bibinfo{number}{8} (\bibinfo{year}{2020}), \bibinfo{pages}{3549--3568}.
\newblock


\bibitem[Guo et~al\mbox{.}(2023)]%
        {guo2023disentangled}
\bibfield{author}{\bibinfo{person}{Xiaobo Guo}, \bibinfo{person}{Shaoshuai Li}, \bibinfo{person}{Naicheng Guo}, \bibinfo{person}{Jiangxia Cao}, \bibinfo{person}{Xiaolei Liu}, \bibinfo{person}{Qiongxu Ma}, \bibinfo{person}{Runsheng Gan}, {and} \bibinfo{person}{Yunan Zhao}.} \bibinfo{year}{2023}\natexlab{}.
\newblock \showarticletitle{Disentangled Representations Learning for Multi-target Cross-domain Recommendation}.
\newblock \bibinfo{journal}{\emph{ACM Transactions on Information Systems}} \bibinfo{volume}{41}, \bibinfo{number}{4} (\bibinfo{year}{2023}), \bibinfo{pages}{1--27}.
\newblock


\bibitem[Han et~al\mbox{.}(2022)]%
        {xing2022streaming}
\bibfield{author}{\bibinfo{person}{Xin Han}, \bibinfo{person}{Ye Zhu}, \bibinfo{person}{Kai~Ming Ting}, \bibinfo{person}{De-Chuan Zhan}, {and} \bibinfo{person}{Gang Li}.} \bibinfo{year}{2022}\natexlab{}.
\newblock \showarticletitle{Streaming Hierarchical Clustering Based on Point-Set Kernel}. In \bibinfo{booktitle}{\emph{Proceedings of the 28th ACM SIGKDD Conference on Knowledge Discovery and Data Mining}} (Washington DC, USA) \emph{(\bibinfo{series}{KDD '22})}. \bibinfo{publisher}{Association for Computing Machinery}, \bibinfo{address}{New York, NY, USA}, \bibinfo{pages}{525–533}.
\newblock
\showISBNx{9781450393850}
\urldef\tempurl%
\url{https://doi.org/10.1145/3534678.3539323}
\showDOI{\tempurl}


\bibitem[Harte et~al\mbox{.}(2023)]%
        {harte2023leveraging}
\bibfield{author}{\bibinfo{person}{Jesse Harte}, \bibinfo{person}{Wouter Zorgdrager}, \bibinfo{person}{Panos Louridas}, \bibinfo{person}{Asterios Katsifodimos}, \bibinfo{person}{Dietmar Jannach}, {and} \bibinfo{person}{Marios Fragkoulis}.} \bibinfo{year}{2023}\natexlab{}.
\newblock \showarticletitle{Leveraging Large Language Models for Sequential Recommendation}. In \bibinfo{booktitle}{\emph{Proceedings of the 17th ACM Conference on Recommender Systems}}. \bibinfo{pages}{1096--1102}.
\newblock


\bibitem[He and McAuley(2016)]%
        {he2016ups}
\bibfield{author}{\bibinfo{person}{Ruining He} {and} \bibinfo{person}{Julian McAuley}.} \bibinfo{year}{2016}\natexlab{}.
\newblock \showarticletitle{Ups and downs: Modeling the visual evolution of fashion trends with one-class collaborative filtering}. In \bibinfo{booktitle}{\emph{proceedings of the 25th international conference on world wide web}}. \bibinfo{pages}{507--517}.
\newblock


\bibitem[He et~al\mbox{.}(2023)]%
        {he2023large}
\bibfield{author}{\bibinfo{person}{Zhankui He}, \bibinfo{person}{Zhouhang Xie}, \bibinfo{person}{Rahul Jha}, \bibinfo{person}{Harald Steck}, \bibinfo{person}{Dawen Liang}, \bibinfo{person}{Yesu Feng}, \bibinfo{person}{Bodhisattwa~Prasad Majumder}, \bibinfo{person}{Nathan Kallus}, {and} \bibinfo{person}{Julian McAuley}.} \bibinfo{year}{2023}\natexlab{}.
\newblock \showarticletitle{Large language models as zero-shot conversational recommenders}. In \bibinfo{booktitle}{\emph{Proceedings of the 32nd ACM international conference on information and knowledge management}}. \bibinfo{pages}{720--730}.
\newblock


\bibitem[Hou et~al\mbox{.}(2023a)]%
        {VQRec}
\bibfield{author}{\bibinfo{person}{Yupeng Hou}, \bibinfo{person}{Zhankui He}, \bibinfo{person}{Julian McAuley}, {and} \bibinfo{person}{Wayne~Xin Zhao}.} \bibinfo{year}{2023}\natexlab{a}.
\newblock \showarticletitle{Learning Vector-Quantized Item Representation for Transferable Sequential Recommenders}. In \bibinfo{booktitle}{\emph{Proceedings of the ACM Web Conference 2023}}. \bibinfo{pages}{1162–1171}.
\newblock


\bibitem[Hou et~al\mbox{.}(2022)]%
        {UniSRec}
\bibfield{author}{\bibinfo{person}{Yupeng Hou}, \bibinfo{person}{Shanlei Mu}, \bibinfo{person}{Wayne~Xin Zhao}, \bibinfo{person}{Yaliang Li}, \bibinfo{person}{Bolin Ding}, {and} \bibinfo{person}{Ji-Rong Wen}.} \bibinfo{year}{2022}\natexlab{}.
\newblock \showarticletitle{Towards Universal Sequence Representation Learning for Recommender Systems}. In \bibinfo{booktitle}{\emph{Proceedings of the 28th ACM SIGKDD Conference on Knowledge Discovery and Data Mining}}. \bibinfo{pages}{585–593}.
\newblock


\bibitem[Hou et~al\mbox{.}(2023b)]%
        {hou2023large}
\bibfield{author}{\bibinfo{person}{Yupeng Hou}, \bibinfo{person}{Junjie Zhang}, \bibinfo{person}{Zihan Lin}, \bibinfo{person}{Hongyu Lu}, \bibinfo{person}{Ruobing Xie}, \bibinfo{person}{Julian McAuley}, {and} \bibinfo{person}{Wayne~Xin Zhao}.} \bibinfo{year}{2023}\natexlab{b}.
\newblock \showarticletitle{Large Language Models are Zero-Shot Rankers for Recommender Systems}.
\newblock \bibinfo{journal}{\emph{arXiv preprint arXiv:2305.08845}} (\bibinfo{year}{2023}).
\newblock


\bibitem[Hu et~al\mbox{.}(2021)]%
        {hu2021lora}
\bibfield{author}{\bibinfo{person}{Edward~J Hu}, \bibinfo{person}{Phillip Wallis}, \bibinfo{person}{Zeyuan Allen-Zhu}, \bibinfo{person}{Yuanzhi Li}, \bibinfo{person}{Shean Wang}, \bibinfo{person}{Lu Wang}, \bibinfo{person}{Weizhu Chen}, {et~al\mbox{.}}} \bibinfo{year}{2021}\natexlab{}.
\newblock \showarticletitle{LoRA: Low-Rank Adaptation of Large Language Models}. In \bibinfo{booktitle}{\emph{International Conference on Learning Representations}}.
\newblock


\bibitem[Huang and Chang(2022)]%
        {huang2022reasoning}
\bibfield{author}{\bibinfo{person}{Jie Huang} {and} \bibinfo{person}{Kevin Chen-Chuan Chang}.} \bibinfo{year}{2022}\natexlab{}.
\newblock \bibinfo{title}{Towards Reasoning in Large Language Models: A Survey}.
\newblock
\newblock
\showeprint[arxiv]{2212.10403}~[cs.CL]


\bibitem[Huang et~al\mbox{.}(2019)]%
        {FiBiNET}
\bibfield{author}{\bibinfo{person}{Tongwen Huang}, \bibinfo{person}{Zhiqi Zhang}, {and} \bibinfo{person}{Junlin Zhang}.} \bibinfo{year}{2019}\natexlab{}.
\newblock \showarticletitle{FiBiNET: Combining Feature Importance and Bilinear Feature Interaction for Click-through Rate Prediction}. In \bibinfo{booktitle}{\emph{Proceedings of the 13th ACM Conference on Recommender Systems}}. \bibinfo{pages}{169–177}.
\newblock


\bibitem[Jacobs et~al\mbox{.}(1991)]%
        {MoE}
\bibfield{author}{\bibinfo{person}{Robert~A. Jacobs}, \bibinfo{person}{Michael~I. Jordan}, \bibinfo{person}{Steven~J. Nowlan}, {and} \bibinfo{person}{Geoffrey~E. Hinton}.} \bibinfo{year}{1991}\natexlab{}.
\newblock \showarticletitle{{Adaptive Mixtures of Local Experts}}.
\newblock \bibinfo{journal}{\emph{Neural Computation}} \bibinfo{volume}{3}, \bibinfo{number}{1} (\bibinfo{date}{03} \bibinfo{year}{1991}), \bibinfo{pages}{79--87}.
\newblock


\bibitem[J\"{a}rvelin and Kek\"{a}l\"{a}inen(2002)]%
        {ndcg}
\bibfield{author}{\bibinfo{person}{Kalervo J\"{a}rvelin} {and} \bibinfo{person}{Jaana Kek\"{a}l\"{a}inen}.} \bibinfo{year}{2002}\natexlab{}.
\newblock \showarticletitle{Cumulated Gain-Based Evaluation of IR Techniques}.
\newblock \bibinfo{journal}{\emph{ACM Trans. Inf. Syst.}} (\bibinfo{year}{2002}), \bibinfo{pages}{422–446}.
\newblock


\bibitem[Ji et~al\mbox{.}(2023a)]%
        {ji2023survey}
\bibfield{author}{\bibinfo{person}{Ziwei Ji}, \bibinfo{person}{Nayeon Lee}, \bibinfo{person}{Rita Frieske}, \bibinfo{person}{Tiezheng Yu}, \bibinfo{person}{Dan Su}, \bibinfo{person}{Yan Xu}, \bibinfo{person}{Etsuko Ishii}, \bibinfo{person}{Ye~Jin Bang}, \bibinfo{person}{Andrea Madotto}, {and} \bibinfo{person}{Pascale Fung}.} \bibinfo{year}{2023}\natexlab{a}.
\newblock \showarticletitle{Survey of hallucination in natural language generation}.
\newblock \bibinfo{journal}{\emph{Comput. Surveys}} \bibinfo{volume}{55}, \bibinfo{number}{12} (\bibinfo{year}{2023}), \bibinfo{pages}{1--38}.
\newblock


\bibitem[Ji et~al\mbox{.}(2023b)]%
        {ji2023teachers}
\bibfield{author}{\bibinfo{person}{Zhong Ji}, \bibinfo{person}{Jingwei Ni}, \bibinfo{person}{Xiyao Liu}, {and} \bibinfo{person}{Yanwei Pang}.} \bibinfo{year}{2023}\natexlab{b}.
\newblock \showarticletitle{Teachers cooperation: team-knowledge distillation for multiple cross-domain few-shot learning}.
\newblock \bibinfo{journal}{\emph{Frontiers of Computer Science}} \bibinfo{volume}{17}, \bibinfo{number}{2} (\bibinfo{year}{2023}), \bibinfo{pages}{172312}.
\newblock


\bibitem[Jiao et~al\mbox{.}(2019)]%
        {jiao2019tinybert}
\bibfield{author}{\bibinfo{person}{Xiaoqi Jiao}, \bibinfo{person}{Yichun Yin}, \bibinfo{person}{Lifeng Shang}, \bibinfo{person}{Xin Jiang}, \bibinfo{person}{Xiao Chen}, \bibinfo{person}{Linlin Li}, \bibinfo{person}{Fang Wang}, {and} \bibinfo{person}{Qun Liu}.} \bibinfo{year}{2019}\natexlab{}.
\newblock \showarticletitle{Tinybert: Distilling bert for natural language understanding}.
\newblock \bibinfo{journal}{\emph{arXiv preprint arXiv:1909.10351}} (\bibinfo{year}{2019}).
\newblock


\bibitem[Kang et~al\mbox{.}(2023)]%
        {kang2023llms}
\bibfield{author}{\bibinfo{person}{Wang-Cheng Kang}, \bibinfo{person}{Jianmo Ni}, \bibinfo{person}{Nikhil Mehta}, \bibinfo{person}{Maheswaran Sathiamoorthy}, \bibinfo{person}{Lichan Hong}, \bibinfo{person}{Ed Chi}, {and} \bibinfo{person}{Derek~Zhiyuan Cheng}.} \bibinfo{year}{2023}\natexlab{}.
\newblock \showarticletitle{Do LLMs Understand User Preferences? Evaluating LLMs On User Rating Prediction}.
\newblock \bibinfo{journal}{\emph{arXiv preprint arXiv:2305.06474}} (\bibinfo{year}{2023}).
\newblock


\bibitem[Kenton and Toutanova(2019)]%
        {bert}
\bibfield{author}{\bibinfo{person}{Jacob Devlin Ming-Wei~Chang Kenton} {and} \bibinfo{person}{Lee~Kristina Toutanova}.} \bibinfo{year}{2019}\natexlab{}.
\newblock \showarticletitle{BERT: Pre-training of Deep Bidirectional Transformers for Language Understanding}. In \bibinfo{booktitle}{\emph{Proceedings of NAACL-HLT}}. \bibinfo{pages}{4171--4186}.
\newblock


\bibitem[Kong et~al\mbox{.}(2021)]%
        {kong2021review}
\bibfield{author}{\bibinfo{person}{Zhiyu Kong}, \bibinfo{person}{Xiaoru Zhang}, {and} \bibinfo{person}{Ruilin Wang}.} \bibinfo{year}{2021}\natexlab{}.
\newblock \showarticletitle{Review of the Research on the Relationship Between Algorithmic News Recommendation and Information Cocoons}. In \bibinfo{booktitle}{\emph{2021 3rd International Conference on Literature, Art and Human Development (ICLAHD 2021)}}. Atlantis Press, \bibinfo{pages}{341--345}.
\newblock


\bibitem[Koren et~al\mbox{.}(2009)]%
        {koren2009matrix}
\bibfield{author}{\bibinfo{person}{Yehuda Koren}, \bibinfo{person}{Robert Bell}, {and} \bibinfo{person}{Chris Volinsky}.} \bibinfo{year}{2009}\natexlab{}.
\newblock \showarticletitle{Matrix factorization techniques for recommender systems}.
\newblock \bibinfo{journal}{\emph{Computer}} \bibinfo{volume}{42}, \bibinfo{number}{8} (\bibinfo{year}{2009}), \bibinfo{pages}{30--37}.
\newblock


\bibitem[Li et~al\mbox{.}(2023a)]%
        {li2023taggpt}
\bibfield{author}{\bibinfo{person}{Chen Li}, \bibinfo{person}{Yixiao Ge}, \bibinfo{person}{Jiayong Mao}, \bibinfo{person}{Dian Li}, {and} \bibinfo{person}{Ying Shan}.} \bibinfo{year}{2023}\natexlab{a}.
\newblock \showarticletitle{TagGPT: Large Language Models are Zero-shot Multimodal Taggers}.
\newblock \bibinfo{journal}{\emph{arXiv preprint arXiv:2304.03022}} (\bibinfo{year}{2023}).
\newblock


\bibitem[Li et~al\mbox{.}(2023b)]%
        {li2023text}
\bibfield{author}{\bibinfo{person}{Jiacheng Li}, \bibinfo{person}{Ming Wang}, \bibinfo{person}{Jin Li}, \bibinfo{person}{Jinmiao Fu}, \bibinfo{person}{Xin Shen}, \bibinfo{person}{Jingbo Shang}, {and} \bibinfo{person}{Julian McAuley}.} \bibinfo{year}{2023}\natexlab{b}.
\newblock \showarticletitle{Text Is All You Need: Learning Language Representations for Sequential Recommendation}.
\newblock \bibinfo{journal}{\emph{arXiv preprint arXiv:2305.13731}} (\bibinfo{year}{2023}).
\newblock


\bibitem[Li et~al\mbox{.}(2023c)]%
        {li2023large}
\bibfield{author}{\bibinfo{person}{Lei Li}, \bibinfo{person}{Yongfeng Zhang}, \bibinfo{person}{Dugang Liu}, {and} \bibinfo{person}{Li Chen}.} \bibinfo{year}{2023}\natexlab{c}.
\newblock \showarticletitle{Large Language Models for Generative Recommendation: A Survey and Visionary Discussions}.
\newblock \bibinfo{journal}{\emph{arXiv preprint arXiv:2309.01157}} (\bibinfo{year}{2023}).
\newblock


\bibitem[Li et~al\mbox{.}(2019)]%
        {Fi-GNN}
\bibfield{author}{\bibinfo{person}{Zekun Li}, \bibinfo{person}{Zeyu Cui}, \bibinfo{person}{Shu Wu}, \bibinfo{person}{Xiaoyu Zhang}, {and} \bibinfo{person}{Liang Wang}.} \bibinfo{year}{2019}\natexlab{}.
\newblock \showarticletitle{Fi-GNN: Modeling Feature Interactions via Graph Neural Networks for CTR Prediction}. In \bibinfo{booktitle}{\emph{Proceedings of the 28th ACM International Conference on Information and Knowledge Management}}. \bibinfo{pages}{539–548}.
\newblock


\bibitem[Lian et~al\mbox{.}(2018)]%
        {xDeepFM}
\bibfield{author}{\bibinfo{person}{Jianxun Lian}, \bibinfo{person}{Xiaohuan Zhou}, \bibinfo{person}{Fuzheng Zhang}, \bibinfo{person}{Zhongxia Chen}, \bibinfo{person}{Xing Xie}, {and} \bibinfo{person}{Guangzhong Sun}.} \bibinfo{year}{2018}\natexlab{}.
\newblock \showarticletitle{XDeepFM: Combining Explicit and Implicit Feature Interactions for Recommender Systems}. In \bibinfo{booktitle}{\emph{Proceedings of the 24th ACM SIGKDD International Conference on Knowledge Discovery and Data Mining}}. \bibinfo{pages}{1754–1763}.
\newblock


\bibitem[Lin and Zhang(2023)]%
        {lin2023sparks}
\bibfield{author}{\bibinfo{person}{Guo Lin} {and} \bibinfo{person}{Yongfeng Zhang}.} \bibinfo{year}{2023}\natexlab{}.
\newblock \showarticletitle{Sparks of Artificial General Recommender (AGR): Early Experiments with ChatGPT}.
\newblock \bibinfo{journal}{\emph{arXiv preprint arXiv:2305.04518}} (\bibinfo{year}{2023}).
\newblock


\bibitem[Lin et~al\mbox{.}(2023a)]%
        {lin2023clickprompt}
\bibfield{author}{\bibinfo{person}{Jianghao Lin}, \bibinfo{person}{Bo Chen}, \bibinfo{person}{Hangyu Wang}, \bibinfo{person}{Yunjia Xi}, \bibinfo{person}{Yanru Qu}, \bibinfo{person}{Xinyi Dai}, \bibinfo{person}{Kangning Zhang}, \bibinfo{person}{Ruiming Tang}, \bibinfo{person}{Yong Yu}, {and} \bibinfo{person}{Weinan Zhang}.} \bibinfo{year}{2023}\natexlab{a}.
\newblock \showarticletitle{ClickPrompt: CTR Models are Strong Prompt Generators for Adapting Language Models to CTR Prediction}.
\newblock \bibinfo{journal}{\emph{arXiv preprint arXiv:2310.09234}} (\bibinfo{year}{2023}).
\newblock


\bibitem[Lin et~al\mbox{.}(2023b)]%
        {lin2023can}
\bibfield{author}{\bibinfo{person}{Jianghao Lin}, \bibinfo{person}{Xinyi Dai}, \bibinfo{person}{Yunjia Xi}, \bibinfo{person}{Weiwen Liu}, \bibinfo{person}{Bo Chen}, \bibinfo{person}{Xiangyang Li}, \bibinfo{person}{Chenxu Zhu}, \bibinfo{person}{Huifeng Guo}, \bibinfo{person}{Yong Yu}, \bibinfo{person}{Ruiming Tang}, {et~al\mbox{.}}} \bibinfo{year}{2023}\natexlab{b}.
\newblock \showarticletitle{How Can Recommender Systems Benefit from Large Language Models: A Survey}.
\newblock \bibinfo{journal}{\emph{arXiv preprint arXiv:2306.05817}} (\bibinfo{year}{2023}).
\newblock


\bibitem[Lin et~al\mbox{.}(2021)]%
        {m6}
\bibfield{author}{\bibinfo{person}{Junyang Lin}, \bibinfo{person}{Rui Men}, \bibinfo{person}{An Yang}, \bibinfo{person}{Chang Zhou}, \bibinfo{person}{Ming Ding}, \bibinfo{person}{Yichang Zhang}, \bibinfo{person}{Peng Wang}, \bibinfo{person}{Ang Wang}, \bibinfo{person}{Le Jiang}, \bibinfo{person}{Xianyan Jia}, {et~al\mbox{.}}} \bibinfo{year}{2021}\natexlab{}.
\newblock \showarticletitle{M6: A chinese multimodal pretrainer}.
\newblock \bibinfo{journal}{\emph{arXiv preprint arXiv:2103.00823}} (\bibinfo{year}{2021}).
\newblock


\bibitem[Lin et~al\mbox{.}(2023c)]%
        {lin2023map}
\bibfield{author}{\bibinfo{person}{Jianghao Lin}, \bibinfo{person}{Yanru Qu}, \bibinfo{person}{Wei Guo}, \bibinfo{person}{Xinyi Dai}, \bibinfo{person}{Ruiming Tang}, \bibinfo{person}{Yong Yu}, {and} \bibinfo{person}{Weinan Zhang}.} \bibinfo{year}{2023}\natexlab{c}.
\newblock \showarticletitle{MAP: A Model-agnostic Pretraining Framework for Click-through Rate Prediction}. In \bibinfo{booktitle}{\emph{Proceedings of the 29th ACM SIGKDD Conference on Knowledge Discovery and Data Mining}}. \bibinfo{pages}{1384--1395}.
\newblock


\bibitem[Lin et~al\mbox{.}(2023d)]%
        {lin2023rella}
\bibfield{author}{\bibinfo{person}{Jianghao Lin}, \bibinfo{person}{Rong Shan}, \bibinfo{person}{Chenxu Zhu}, \bibinfo{person}{Kounianhua Du}, \bibinfo{person}{Bo Chen}, \bibinfo{person}{Shigang Quan}, \bibinfo{person}{Ruiming Tang}, \bibinfo{person}{Yong Yu}, {and} \bibinfo{person}{Weinan Zhang}.} \bibinfo{year}{2023}\natexlab{d}.
\newblock \showarticletitle{ReLLa: Retrieval-enhanced Large Language Models for Lifelong Sequential Behavior Comprehension in Recommendation}.
\newblock \bibinfo{journal}{\emph{arXiv preprint arXiv:2308.11131}} (\bibinfo{year}{2023}).
\newblock


\bibitem[Liu et~al\mbox{.}(2023a)]%
        {liu2023chatgpt}
\bibfield{author}{\bibinfo{person}{Junling Liu}, \bibinfo{person}{Chao Liu}, \bibinfo{person}{Renjie Lv}, \bibinfo{person}{Kang Zhou}, {and} \bibinfo{person}{Yan Zhang}.} \bibinfo{year}{2023}\natexlab{a}.
\newblock \showarticletitle{Is ChatGPT a Good Recommender? A Preliminary Study}.
\newblock \bibinfo{journal}{\emph{arXiv preprint arXiv:2304.10149}} (\bibinfo{year}{2023}).
\newblock


\bibitem[Liu et~al\mbox{.}(2023b)]%
        {liu2023pre}
\bibfield{author}{\bibinfo{person}{Peng Liu}, \bibinfo{person}{Lemei Zhang}, {and} \bibinfo{person}{Jon~Atle Gulla}.} \bibinfo{year}{2023}\natexlab{b}.
\newblock \showarticletitle{Pre-train, prompt and recommendation: A comprehensive survey of language modelling paradigm adaptations in recommender systems}.
\newblock \bibinfo{journal}{\emph{arXiv preprint arXiv:2302.03735}} (\bibinfo{year}{2023}).
\newblock


\bibitem[Liu et~al\mbox{.}(2022)]%
        {liu2022neural}
\bibfield{author}{\bibinfo{person}{Weiwen Liu}, \bibinfo{person}{Yunjia Xi}, \bibinfo{person}{Jiarui Qin}, \bibinfo{person}{Fei Sun}, \bibinfo{person}{Bo Chen}, \bibinfo{person}{Weinan Zhang}, \bibinfo{person}{Rui Zhang}, {and} \bibinfo{person}{Ruiming Tang}.} \bibinfo{year}{2022}\natexlab{}.
\newblock \showarticletitle{Neural Re-ranking in Multi-stage Recommender Systems: A Review}.
\newblock \bibinfo{journal}{\emph{arXiv preprint arXiv:2202.06602}} (\bibinfo{year}{2022}).
\newblock


\bibitem[Liu et~al\mbox{.}(2023c)]%
        {liu2023joint}
\bibfield{author}{\bibinfo{person}{Xiaojian Liu}, \bibinfo{person}{Yi Zhu}, {and} \bibinfo{person}{Xindong Wu}.} \bibinfo{year}{2023}\natexlab{c}.
\newblock \showarticletitle{Joint user profiling with hierarchical attention networks}.
\newblock \bibinfo{journal}{\emph{Frontiers of Computer Science}} \bibinfo{volume}{17}, \bibinfo{number}{3} (\bibinfo{year}{2023}), \bibinfo{pages}{173608}.
\newblock


\bibitem[Luo et~al\mbox{.}(2023)]%
        {luo2023unlocking}
\bibfield{author}{\bibinfo{person}{Yucong Luo}, \bibinfo{person}{Mingyue Cheng}, \bibinfo{person}{Hao Zhang}, \bibinfo{person}{Junyu Lu}, {and} \bibinfo{person}{Enhong Chen}.} \bibinfo{year}{2023}\natexlab{}.
\newblock \showarticletitle{Unlocking the Potential of Large Language Models for Explainable Recommendations}.
\newblock \bibinfo{journal}{\emph{arXiv preprint arXiv:2312.15661}} (\bibinfo{year}{2023}).
\newblock


\bibitem[Lyu et~al\mbox{.}(2023)]%
        {lyu2023llm}
\bibfield{author}{\bibinfo{person}{Hanjia Lyu}, \bibinfo{person}{Song Jiang}, \bibinfo{person}{Hanqing Zeng}, \bibinfo{person}{Yinglong Xia}, {and} \bibinfo{person}{Jiebo Luo}.} \bibinfo{year}{2023}\natexlab{}.
\newblock \showarticletitle{Llm-rec: Personalized recommendation via prompting large language models}.
\newblock \bibinfo{journal}{\emph{arXiv preprint arXiv:2307.15780}} (\bibinfo{year}{2023}).
\newblock


\bibitem[Murtagh and Contreras(2012)]%
        {murtagh2012algorithms}
\bibfield{author}{\bibinfo{person}{Fionn Murtagh} {and} \bibinfo{person}{Pedro Contreras}.} \bibinfo{year}{2012}\natexlab{}.
\newblock \showarticletitle{Algorithms for hierarchical clustering: an overview}.
\newblock \bibinfo{journal}{\emph{Wiley Interdisciplinary Reviews: Data Mining and Knowledge Discovery}} \bibinfo{volume}{2}, \bibinfo{number}{1} (\bibinfo{year}{2012}), \bibinfo{pages}{86--97}.
\newblock


\bibitem[Mysore et~al\mbox{.}(2023)]%
        {mysore2023large}
\bibfield{author}{\bibinfo{person}{Sheshera Mysore}, \bibinfo{person}{Andrew McCallum}, {and} \bibinfo{person}{Hamed Zamani}.} \bibinfo{year}{2023}\natexlab{}.
\newblock \showarticletitle{Large Language Model Augmented Narrative Driven Recommendations}.
\newblock \bibinfo{journal}{\emph{arXiv preprint arXiv:2306.02250}} (\bibinfo{year}{2023}).
\newblock


\bibitem[Ni et~al\mbox{.}(2019)]%
        {ni2019justifying}
\bibfield{author}{\bibinfo{person}{Jianmo Ni}, \bibinfo{person}{Jiacheng Li}, {and} \bibinfo{person}{Julian McAuley}.} \bibinfo{year}{2019}\natexlab{}.
\newblock \showarticletitle{Justifying recommendations using distantly-labeled reviews and fine-grained aspects}. In \bibinfo{booktitle}{\emph{Proceedings of the 2019 conference on empirical methods in natural language processing and the 9th international joint conference on natural language processing (EMNLP-IJCNLP)}}. \bibinfo{pages}{188--197}.
\newblock


\bibitem[OpenAI(2023)]%
        {gpt4}
\bibfield{author}{\bibinfo{person}{OpenAI}.} \bibinfo{year}{2023}\natexlab{}.
\newblock \showarticletitle{{GPT-4} Technical Report}.
\newblock \bibinfo{journal}{\emph{CoRR}}  \bibinfo{volume}{abs/2303.08774} (\bibinfo{year}{2023}).
\newblock
\urldef\tempurl%
\url{https://doi.org/10.48550/arXiv.2303.08774}
\showURL{%
\tempurl}


\bibitem[Ouyang et~al\mbox{.}(2022)]%
        {ouyang2022training}
\bibfield{author}{\bibinfo{person}{Long Ouyang}, \bibinfo{person}{Jeffrey Wu}, \bibinfo{person}{Xu Jiang}, \bibinfo{person}{Diogo Almeida}, \bibinfo{person}{Carroll Wainwright}, \bibinfo{person}{Pamela Mishkin}, \bibinfo{person}{Chong Zhang}, \bibinfo{person}{Sandhini Agarwal}, \bibinfo{person}{Katarina Slama}, \bibinfo{person}{Alex Ray}, {et~al\mbox{.}}} \bibinfo{year}{2022}\natexlab{}.
\newblock \showarticletitle{Training language models to follow instructions with human feedback}.
\newblock \bibinfo{journal}{\emph{Advances in Neural Information Processing Systems}}  \bibinfo{volume}{35} (\bibinfo{year}{2022}), \bibinfo{pages}{27730--27744}.
\newblock


\bibitem[Pan et~al\mbox{.}(2020)]%
        {pan2020privacy}
\bibfield{author}{\bibinfo{person}{Xudong Pan}, \bibinfo{person}{Mi Zhang}, \bibinfo{person}{Shouling Ji}, {and} \bibinfo{person}{Min Yang}.} \bibinfo{year}{2020}\natexlab{}.
\newblock \showarticletitle{Privacy risks of general-purpose language models}. In \bibinfo{booktitle}{\emph{2020 IEEE Symposium on Security and Privacy (SP)}}. IEEE, \bibinfo{pages}{1314--1331}.
\newblock


\bibitem[Pang et~al\mbox{.}(2020)]%
        {pang2020setrank}
\bibfield{author}{\bibinfo{person}{Liang Pang}, \bibinfo{person}{Jun Xu}, \bibinfo{person}{Qingyao Ai}, \bibinfo{person}{Yanyan Lan}, \bibinfo{person}{Xueqi Cheng}, {and} \bibinfo{person}{Jirong Wen}.} \bibinfo{year}{2020}\natexlab{}.
\newblock \showarticletitle{Setrank: Learning a permutation-invariant ranking model for information retrieval}. In \bibinfo{booktitle}{\emph{Proceedings of the 43rd International ACM SIGIR Conference on Research and Development in Information Retrieval}}. \bibinfo{pages}{499--508}.
\newblock


\bibitem[Pei et~al\mbox{.}(2019)]%
        {pei2019personalized}
\bibfield{author}{\bibinfo{person}{Changhua Pei}, \bibinfo{person}{Yi Zhang}, \bibinfo{person}{Yongfeng Zhang}, \bibinfo{person}{Fei Sun}, \bibinfo{person}{Xiao Lin}, \bibinfo{person}{Hanxiao Sun}, \bibinfo{person}{Jian Wu}, \bibinfo{person}{Peng Jiang}, \bibinfo{person}{Junfeng Ge}, \bibinfo{person}{Wenwu Ou}, {et~al\mbox{.}}} \bibinfo{year}{2019}\natexlab{}.
\newblock \showarticletitle{Personalized re-ranking for recommendation}. In \bibinfo{booktitle}{\emph{Proceedings of the 13th ACM conference on recommender systems}}. \bibinfo{pages}{3--11}.
\newblock


\bibitem[Press et~al\mbox{.}(2022)]%
        {press2022measuring}
\bibfield{author}{\bibinfo{person}{Ofir Press}, \bibinfo{person}{Muru Zhang}, \bibinfo{person}{Sewon Min}, \bibinfo{person}{Ludwig Schmidt}, \bibinfo{person}{Noah~A. Smith}, {and} \bibinfo{person}{Mike Lewis}.} \bibinfo{year}{2022}\natexlab{}.
\newblock \bibinfo{title}{Measuring and Narrowing the Compositionality Gap in Language Models}.
\newblock
\newblock
\showeprint[arxiv]{2210.03350}~[cs.CL]


\bibitem[Qiao et~al\mbox{.}(2023)]%
        {qiao2023reasoning}
\bibfield{author}{\bibinfo{person}{Shuofei Qiao}, \bibinfo{person}{Yixin Ou}, \bibinfo{person}{Ningyu Zhang}, \bibinfo{person}{Xiang Chen}, \bibinfo{person}{Yunzhi Yao}, \bibinfo{person}{Shumin Deng}, \bibinfo{person}{Chuanqi Tan}, \bibinfo{person}{Fei Huang}, {and} \bibinfo{person}{Huajun Chen}.} \bibinfo{year}{2023}\natexlab{}.
\newblock \bibinfo{title}{Reasoning with Language Model Prompting: A Survey}.
\newblock
\newblock
\showeprint[arxiv]{2212.09597}~[cs.CL]


\bibitem[Qiu et~al\mbox{.}(2021)]%
        {ubert}
\bibfield{author}{\bibinfo{person}{Zhaopeng Qiu}, \bibinfo{person}{Xian Wu}, \bibinfo{person}{Jingyue Gao}, {and} \bibinfo{person}{Wei Fan}.} \bibinfo{year}{2021}\natexlab{}.
\newblock \showarticletitle{U-BERT: Pre-training user representations for improved recommendation}. In \bibinfo{booktitle}{\emph{Proceedings of the AAAI Conference on Artificial Intelligence}}, Vol.~\bibinfo{volume}{35}. \bibinfo{pages}{4320--4327}.
\newblock


\bibitem[Radford et~al\mbox{.}(2019)]%
        {gpt2}
\bibfield{author}{\bibinfo{person}{Alec Radford}, \bibinfo{person}{Jeffrey Wu}, \bibinfo{person}{Rewon Child}, \bibinfo{person}{David Luan}, \bibinfo{person}{Dario Amodei}, \bibinfo{person}{Ilya Sutskever}, {et~al\mbox{.}}} \bibinfo{year}{2019}\natexlab{}.
\newblock \showarticletitle{Language models are unsupervised multitask learners}.
\newblock \bibinfo{journal}{\emph{OpenAI blog}} \bibinfo{volume}{1}, \bibinfo{number}{8} (\bibinfo{year}{2019}), \bibinfo{pages}{9}.
\newblock


\bibitem[Raffel et~al\mbox{.}(2020)]%
        {t5}
\bibfield{author}{\bibinfo{person}{Colin Raffel}, \bibinfo{person}{Noam Shazeer}, \bibinfo{person}{Adam Roberts}, \bibinfo{person}{Katherine Lee}, \bibinfo{person}{Sharan Narang}, \bibinfo{person}{Michael Matena}, \bibinfo{person}{Yanqi Zhou}, \bibinfo{person}{Wei Li}, {and} \bibinfo{person}{Peter~J Liu}.} \bibinfo{year}{2020}\natexlab{}.
\newblock \showarticletitle{Exploring the limits of transfer learning with a unified text-to-text transformer}.
\newblock \bibinfo{journal}{\emph{The Journal of Machine Learning Research}} \bibinfo{volume}{21}, \bibinfo{number}{1} (\bibinfo{year}{2020}), \bibinfo{pages}{5485--5551}.
\newblock


\bibitem[Rendle(2010)]%
        {rendle2010factorization}
\bibfield{author}{\bibinfo{person}{Steffen Rendle}.} \bibinfo{year}{2010}\natexlab{}.
\newblock \showarticletitle{Factorization machines}. In \bibinfo{booktitle}{\emph{2010 IEEE International conference on data mining}}. IEEE, \bibinfo{pages}{995--1000}.
\newblock


\bibitem[Sanner et~al\mbox{.}(2023)]%
        {sanner2023large}
\bibfield{author}{\bibinfo{person}{Scott Sanner}, \bibinfo{person}{Krisztian Balog}, \bibinfo{person}{Filip Radlinski}, \bibinfo{person}{Ben Wedin}, {and} \bibinfo{person}{Lucas Dixon}.} \bibinfo{year}{2023}\natexlab{}.
\newblock \showarticletitle{Large language models are competitive near cold-start recommenders for language-and item-based preferences}. In \bibinfo{booktitle}{\emph{Proceedings of the 17th ACM conference on recommender systems}}. \bibinfo{pages}{890--896}.
\newblock


\bibitem[Sheng et~al\mbox{.}(2021)]%
        {star}
\bibfield{author}{\bibinfo{person}{Xiang-Rong Sheng}, \bibinfo{person}{Liqin Zhao}, \bibinfo{person}{Guorui Zhou}, \bibinfo{person}{Xinyao Ding}, \bibinfo{person}{Binding Dai}, \bibinfo{person}{Qiang Luo}, \bibinfo{person}{Siran Yang}, \bibinfo{person}{Jingshan Lv}, \bibinfo{person}{Chi Zhang}, \bibinfo{person}{Hongbo Deng}, {et~al\mbox{.}}} \bibinfo{year}{2021}\natexlab{}.
\newblock \showarticletitle{One model to serve all: Star topology adaptive recommender for multi-domain ctr prediction}. In \bibinfo{booktitle}{\emph{Proceedings of the 30th ACM International Conference on Information \& Knowledge Management}}. \bibinfo{pages}{4104--4113}.
\newblock


\bibitem[Song et~al\mbox{.}(2019)]%
        {AutoInt}
\bibfield{author}{\bibinfo{person}{Weiping Song}, \bibinfo{person}{Chence Shi}, \bibinfo{person}{Zhiping Xiao}, \bibinfo{person}{Zhijian Duan}, \bibinfo{person}{Yewen Xu}, \bibinfo{person}{Ming Zhang}, {and} \bibinfo{person}{Jian Tang}.} \bibinfo{year}{2019}\natexlab{}.
\newblock \showarticletitle{AutoInt: Automatic Feature Interaction Learning via Self-Attentive Neural Networks}. In \bibinfo{booktitle}{\emph{Proceedings of the 28th ACM International Conference on Information and Knowledge Management}}. \bibinfo{pages}{1161–1170}.
\newblock


\bibitem[Touvron et~al\mbox{.}(2023)]%
        {llama}
\bibfield{author}{\bibinfo{person}{Hugo Touvron}, \bibinfo{person}{Thibaut Lavril}, \bibinfo{person}{Gautier Izacard}, \bibinfo{person}{Xavier Martinet}, \bibinfo{person}{Marie-Anne Lachaux}, \bibinfo{person}{Timoth{\'e}e Lacroix}, \bibinfo{person}{Baptiste Rozi{\`e}re}, \bibinfo{person}{Naman Goyal}, \bibinfo{person}{Eric Hambro}, \bibinfo{person}{Faisal Azhar}, {et~al\mbox{.}}} \bibinfo{year}{2023}\natexlab{}.
\newblock \showarticletitle{Llama: Open and efficient foundation language models}.
\newblock \bibinfo{journal}{\emph{arXiv preprint arXiv:2302.13971}} (\bibinfo{year}{2023}).
\newblock


\bibitem[Van~den Oord et~al\mbox{.}(2013)]%
        {van2013deep}
\bibfield{author}{\bibinfo{person}{Aaron Van~den Oord}, \bibinfo{person}{Sander Dieleman}, {and} \bibinfo{person}{Benjamin Schrauwen}.} \bibinfo{year}{2013}\natexlab{}.
\newblock \showarticletitle{Deep content-based music recommendation}.
\newblock \bibinfo{journal}{\emph{Advances in neural information processing systems}}  \bibinfo{volume}{26} (\bibinfo{year}{2013}).
\newblock


\bibitem[Wang et~al\mbox{.}(2023)]%
        {wang2023flip}
\bibfield{author}{\bibinfo{person}{Hangyu Wang}, \bibinfo{person}{Jianghao Lin}, \bibinfo{person}{Xiangyang Li}, \bibinfo{person}{Bo Chen}, \bibinfo{person}{Chenxu Zhu}, \bibinfo{person}{Ruiming Tang}, \bibinfo{person}{Weinan Zhang}, {and} \bibinfo{person}{Yong Yu}.} \bibinfo{year}{2023}\natexlab{}.
\newblock \showarticletitle{FLIP: Towards Fine-grained Alignment between ID-based Models and Pretrained Language Models for CTR Prediction}.
\newblock \bibinfo{journal}{\emph{arXiv e-prints}} (\bibinfo{year}{2023}), \bibinfo{pages}{arXiv--2310}.
\newblock


\bibitem[Wang et~al\mbox{.}(2019a)]%
        {wang2019multi}
\bibfield{author}{\bibinfo{person}{Hongwei Wang}, \bibinfo{person}{Fuzheng Zhang}, \bibinfo{person}{Miao Zhao}, \bibinfo{person}{Wenjie Li}, \bibinfo{person}{Xing Xie}, {and} \bibinfo{person}{Minyi Guo}.} \bibinfo{year}{2019}\natexlab{a}.
\newblock \showarticletitle{Multi-task feature learning for knowledge graph enhanced recommendation}. In \bibinfo{booktitle}{\emph{The world wide web conference}}. \bibinfo{pages}{2000--2010}.
\newblock


\bibitem[Wang et~al\mbox{.}(2019b)]%
        {wang2019knowledge}
\bibfield{author}{\bibinfo{person}{Hongwei Wang}, \bibinfo{person}{Miao Zhao}, \bibinfo{person}{Xing Xie}, \bibinfo{person}{Wenjie Li}, {and} \bibinfo{person}{Minyi Guo}.} \bibinfo{year}{2019}\natexlab{b}.
\newblock \showarticletitle{Knowledge graph convolutional networks for recommender systems}. In \bibinfo{booktitle}{\emph{The world wide web conference}}. \bibinfo{pages}{3307--3313}.
\newblock


\bibitem[Wang and Lim(2023)]%
        {nir}
\bibfield{author}{\bibinfo{person}{Lei Wang} {and} \bibinfo{person}{Ee-Peng Lim}.} \bibinfo{year}{2023}\natexlab{}.
\newblock \showarticletitle{Zero-Shot Next-Item Recommendation using Large Pretrained Language Models}.
\newblock \bibinfo{journal}{\emph{arXiv preprint arXiv:2304.03153}} (\bibinfo{year}{2023}).
\newblock


\bibitem[Wang et~al\mbox{.}(2017)]%
        {DCN}
\bibfield{author}{\bibinfo{person}{Ruoxi Wang}, \bibinfo{person}{Bin Fu}, \bibinfo{person}{Gang Fu}, {and} \bibinfo{person}{Mingliang Wang}.} \bibinfo{year}{2017}\natexlab{}.
\newblock \showarticletitle{Deep \& Cross Network for Ad Click Predictions}. In \bibinfo{booktitle}{\emph{Proceedings of the ADKDD'17}}. Article \bibinfo{articleno}{12}, \bibinfo{numpages}{7}~pages.
\newblock


\bibitem[Wang et~al\mbox{.}(2021)]%
        {DCNv2}
\bibfield{author}{\bibinfo{person}{Ruoxi Wang}, \bibinfo{person}{Rakesh Shivanna}, \bibinfo{person}{Derek Cheng}, \bibinfo{person}{Sagar Jain}, \bibinfo{person}{Dong Lin}, \bibinfo{person}{Lichan Hong}, {and} \bibinfo{person}{Ed Chi}.} \bibinfo{year}{2021}\natexlab{}.
\newblock \showarticletitle{DCN V2: Improved Deep \& Cross Network and Practical Lessons for Web-Scale Learning to Rank Systems}. In \bibinfo{booktitle}{\emph{Proceedings of the Web Conference 2021}}. \bibinfo{pages}{1785–1797}.
\newblock


\bibitem[Wei et~al\mbox{.}(2022)]%
        {wei2022chain}
\bibfield{author}{\bibinfo{person}{Jason Wei}, \bibinfo{person}{Xuezhi Wang}, \bibinfo{person}{Dale Schuurmans}, \bibinfo{person}{Maarten Bosma}, \bibinfo{person}{Ed Chi}, \bibinfo{person}{Quoc Le}, {and} \bibinfo{person}{Denny Zhou}.} \bibinfo{year}{2022}\natexlab{}.
\newblock \showarticletitle{Chain of thought prompting elicits reasoning in large language models}.
\newblock \bibinfo{journal}{\emph{arXiv preprint arXiv:2201.11903}} (\bibinfo{year}{2022}).
\newblock


\bibitem[Weidinger et~al\mbox{.}(2021)]%
        {weidinger2021ethical}
\bibfield{author}{\bibinfo{person}{Laura Weidinger}, \bibinfo{person}{John Mellor}, \bibinfo{person}{Maribeth Rauh}, \bibinfo{person}{Conor Griffin}, \bibinfo{person}{Jonathan Uesato}, \bibinfo{person}{Po-Sen Huang}, \bibinfo{person}{Myra Cheng}, \bibinfo{person}{Mia Glaese}, \bibinfo{person}{Borja Balle}, \bibinfo{person}{Atoosa Kasirzadeh}, {et~al\mbox{.}}} \bibinfo{year}{2021}\natexlab{}.
\newblock \showarticletitle{Ethical and social risks of harm from language models}.
\newblock \bibinfo{journal}{\emph{arXiv preprint arXiv:2112.04359}} (\bibinfo{year}{2021}).
\newblock


\bibitem[Wu et~al\mbox{.}(2023)]%
        {wu2023survey}
\bibfield{author}{\bibinfo{person}{Likang Wu}, \bibinfo{person}{Zhi Zheng}, \bibinfo{person}{Zhaopeng Qiu}, \bibinfo{person}{Hao Wang}, \bibinfo{person}{Hongchao Gu}, \bibinfo{person}{Tingjia Shen}, \bibinfo{person}{Chuan Qin}, \bibinfo{person}{Chen Zhu}, \bibinfo{person}{Hengshu Zhu}, \bibinfo{person}{Qi Liu}, {et~al\mbox{.}}} \bibinfo{year}{2023}\natexlab{}.
\newblock \showarticletitle{A Survey on Large Language Models for Recommendation}.
\newblock \bibinfo{journal}{\emph{arXiv preprint arXiv:2305.19860}} (\bibinfo{year}{2023}).
\newblock


\bibitem[Xi et~al\mbox{.}(2023)]%
        {DIR}
\bibfield{author}{\bibinfo{person}{Yunjia Xi}, \bibinfo{person}{Weiwen Liu}, \bibinfo{person}{Yang Wang}, \bibinfo{person}{Ruiming Tang}, \bibinfo{person}{Weinan Zhang}, \bibinfo{person}{Yue Zhu}, \bibinfo{person}{Rui Zhang}, {and} \bibinfo{person}{Yong Yu}.} \bibinfo{year}{2023}\natexlab{}.
\newblock \showarticletitle{On-device Integrated Re-ranking with Heterogeneous Behavior Modeling}. In \bibinfo{booktitle}{\emph{Proceedings of the 29th ACM SIGKDD Conference on Knowledge Discovery and Data Mining}} (, Long Beach, CA, USA,) \emph{(\bibinfo{series}{KDD '23})}. \bibinfo{publisher}{Association for Computing Machinery}, \bibinfo{address}{New York, NY, USA}, \bibinfo{pages}{5225–5236}.
\newblock
\showISBNx{9798400701030}
\urldef\tempurl%
\url{https://doi.org/10.1145/3580305.3599878}
\showDOI{\tempurl}


\bibitem[Xi et~al\mbox{.}(2022)]%
        {xi2022multi}
\bibfield{author}{\bibinfo{person}{Yunjia Xi}, \bibinfo{person}{Weiwen Liu}, \bibinfo{person}{Jieming Zhu}, \bibinfo{person}{Xilong Zhao}, \bibinfo{person}{Xinyi Dai}, \bibinfo{person}{Ruiming Tang}, \bibinfo{person}{Weinan Zhang}, \bibinfo{person}{Rui Zhang}, {and} \bibinfo{person}{Yong Yu}.} \bibinfo{year}{2022}\natexlab{}.
\newblock \showarticletitle{Multi-Level Interaction Reranking with User Behavior History}. In \bibinfo{booktitle}{\emph{Proceedings of the 45th International ACM SIGIR Conference on Research and Development in Information Retrieval}}.
\newblock


\bibitem[Yu et~al\mbox{.}(2023)]%
        {yu2023self}
\bibfield{author}{\bibinfo{person}{Junliang Yu}, \bibinfo{person}{Hongzhi Yin}, \bibinfo{person}{Xin Xia}, \bibinfo{person}{Tong Chen}, \bibinfo{person}{Jundong Li}, {and} \bibinfo{person}{Zi Huang}.} \bibinfo{year}{2023}\natexlab{}.
\newblock \showarticletitle{Self-supervised learning for recommender systems: A survey}.
\newblock \bibinfo{journal}{\emph{IEEE Transactions on Knowledge and Data Engineering}} (\bibinfo{year}{2023}).
\newblock


\bibitem[Yue et~al\mbox{.}(2007)]%
        {yue2007support}
\bibfield{author}{\bibinfo{person}{Yisong Yue}, \bibinfo{person}{Thomas Finley}, \bibinfo{person}{Filip Radlinski}, {and} \bibinfo{person}{Thorsten Joachims}.} \bibinfo{year}{2007}\natexlab{}.
\newblock \showarticletitle{A support vector method for optimizing average precision}. In \bibinfo{booktitle}{\emph{Proceedings of the 30th annual international ACM SIGIR conference on Research and development in information retrieval}}. \bibinfo{pages}{271--278}.
\newblock


\bibitem[Zeng et~al\mbox{.}(2021)]%
        {zeng2021pangu}
\bibfield{author}{\bibinfo{person}{Wei Zeng}, \bibinfo{person}{Xiaozhe Ren}, \bibinfo{person}{Teng Su}, \bibinfo{person}{Hui Wang}, \bibinfo{person}{Yi Liao}, \bibinfo{person}{Zhiwei Wang}, \bibinfo{person}{Xin Jiang}, \bibinfo{person}{ZhenZhang Yang}, \bibinfo{person}{Kaisheng Wang}, \bibinfo{person}{Xiaoda Zhang}, {et~al\mbox{.}}} \bibinfo{year}{2021}\natexlab{}.
\newblock \showarticletitle{Pangu-$\alpha$: Large-scale autoregressive pretrained Chinese language models with auto-parallel computation}.
\newblock \bibinfo{journal}{\emph{arXiv preprint arXiv:2104.12369}} (\bibinfo{year}{2021}).
\newblock


\bibitem[Zhang et~al\mbox{.}(2014)]%
        {zhang2014addressing}
\bibfield{author}{\bibinfo{person}{Mi Zhang}, \bibinfo{person}{Jie Tang}, \bibinfo{person}{Xuchen Zhang}, {and} \bibinfo{person}{Xiangyang Xue}.} \bibinfo{year}{2014}\natexlab{}.
\newblock \showarticletitle{Addressing Cold Start in Recommender Systems: A Semi-Supervised Co-Training Algorithm}. In \bibinfo{booktitle}{\emph{Proceedings of the 37th International ACM SIGIR Conference on Research \& Development in Information Retrieval}} \emph{(\bibinfo{series}{SIGIR '14})}. \bibinfo{pages}{73–82}.
\newblock


\bibitem[Zhang et~al\mbox{.}(2021)]%
        {zhang2021language}
\bibfield{author}{\bibinfo{person}{Yuhui Zhang}, \bibinfo{person}{HAO DING}, \bibinfo{person}{Zeren Shui}, \bibinfo{person}{Yifei Ma}, \bibinfo{person}{James Zou}, \bibinfo{person}{Anoop Deoras}, {and} \bibinfo{person}{Hao Wang}.} \bibinfo{year}{2021}\natexlab{}.
\newblock \showarticletitle{Language Models as Recommender Systems: Evaluations and Limitations}. In \bibinfo{booktitle}{\emph{I (Still) Can't Believe It's Not Better! NeurIPS 2021 Workshop}}.
\newblock


\bibitem[Zhao et~al\mbox{.}(2023)]%
        {zhao2023survey}
\bibfield{author}{\bibinfo{person}{Wayne~Xin Zhao}, \bibinfo{person}{Kun Zhou}, \bibinfo{person}{Junyi Li}, \bibinfo{person}{Tianyi Tang}, \bibinfo{person}{Xiaolei Wang}, \bibinfo{person}{Yupeng Hou}, \bibinfo{person}{Yingqian Min}, \bibinfo{person}{Beichen Zhang}, \bibinfo{person}{Junjie Zhang}, \bibinfo{person}{Zican Dong}, {et~al\mbox{.}}} \bibinfo{year}{2023}\natexlab{}.
\newblock \showarticletitle{A survey of large language models}.
\newblock \bibinfo{journal}{\emph{arXiv preprint arXiv:2303.18223}} (\bibinfo{year}{2023}).
\newblock


\bibitem[Zhou et~al\mbox{.}(2023)]%
        {zhou2023leasttomost}
\bibfield{author}{\bibinfo{person}{Denny Zhou}, \bibinfo{person}{Nathanael Schärli}, \bibinfo{person}{Le Hou}, \bibinfo{person}{Jason Wei}, \bibinfo{person}{Nathan Scales}, \bibinfo{person}{Xuezhi Wang}, \bibinfo{person}{Dale Schuurmans}, \bibinfo{person}{Claire Cui}, \bibinfo{person}{Olivier Bousquet}, \bibinfo{person}{Quoc Le}, {and} \bibinfo{person}{Ed Chi}.} \bibinfo{year}{2023}\natexlab{}.
\newblock \showarticletitle{Least-to-Most Prompting Enables Complex Reasoning in Large Language Models}. In \bibinfo{booktitle}{\emph{The Eleventh International Conference on Learning Representations}}.
\newblock


\bibitem[Zhou et~al\mbox{.}(2019)]%
        {DIEN}
\bibfield{author}{\bibinfo{person}{Guorui Zhou}, \bibinfo{person}{Na Mou}, \bibinfo{person}{Ying Fan}, \bibinfo{person}{Qi Pi}, \bibinfo{person}{Weijie Bian}, \bibinfo{person}{Chang Zhou}, \bibinfo{person}{Xiaoqiang Zhu}, {and} \bibinfo{person}{Kun Gai}.} \bibinfo{year}{2019}\natexlab{}.
\newblock \showarticletitle{Deep Interest Evolution Network for Click-through Rate Prediction} \emph{(\bibinfo{series}{AAAI'19})}. Article \bibinfo{articleno}{729}, \bibinfo{numpages}{8}~pages.
\newblock


\bibitem[Zhou et~al\mbox{.}(2018)]%
        {DIN}
\bibfield{author}{\bibinfo{person}{Guorui Zhou}, \bibinfo{person}{Xiaoqiang Zhu}, \bibinfo{person}{Chenru Song}, \bibinfo{person}{Ying Fan}, \bibinfo{person}{Han Zhu}, \bibinfo{person}{Xiao Ma}, \bibinfo{person}{Yanghui Yan}, \bibinfo{person}{Junqi Jin}, \bibinfo{person}{Han Li}, {and} \bibinfo{person}{Kun Gai}.} \bibinfo{year}{2018}\natexlab{}.
\newblock \showarticletitle{Deep Interest Network for Click-Through Rate Prediction}. In \bibinfo{booktitle}{\emph{Proceedings of the 24th ACM SIGKDD International Conference on Knowledge Discovery and Data Mining}}. \bibinfo{pages}{1059–1068}.
\newblock


\bibitem[Zhu et~al\mbox{.}(2023)]%
        {zhu2023large}
\bibfield{author}{\bibinfo{person}{Yutao Zhu}, \bibinfo{person}{Huaying Yuan}, \bibinfo{person}{Shuting Wang}, \bibinfo{person}{Jiongnan Liu}, \bibinfo{person}{Wenhan Liu}, \bibinfo{person}{Chenlong Deng}, \bibinfo{person}{Zhicheng Dou}, {and} \bibinfo{person}{Ji-Rong Wen}.} \bibinfo{year}{2023}\natexlab{}.
\newblock \showarticletitle{Large language models for information retrieval: A survey}.
\newblock \bibinfo{journal}{\emph{arXiv preprint arXiv:2308.07107}} (\bibinfo{year}{2023}).
\newblock


\end{thebibliography}

\end{document}